\documentclass[fleqn,usenatbib]{mnras}

\usepackage{newtxtext,newtxmath}
\usepackage[T1]{fontenc}

\DeclareRobustCommand{\VAN}[3]{#2}
\let\VANthebibliography\thebibliography
\def\thebibliography{\DeclareRobustCommand{\VAN}[3]{##3}\VANthebibliography}

\usepackage[dvipdfmx]{graphicx}	% Including figure files
\usepackage{amsmath}	% Advanced maths commands
\usepackage{bm}
\usepackage{here}
\usepackage{newtxtext,newtxmath}
\newcommand{\beq}{\begin{equation}}
\newcommand{\beqa}{\begin{eqnarray}}
\newcommand{\eeq}{\end{equation}}
\newcommand{\eeqa}{\end{eqnarray}}

\newcommand{\simgt}{\lower.5ex\hbox{$\; \buildrel > \over \sim \;$}}

\newcommand{\mpch}{h^{-1}{\rm Mpc}}
\newcommand{\msunh}{h^{-1}M_\odot}

\def\avrg#1{\left\langle #1 \right\rangle}

\usepackage{amsmath} % or simply amstext
\newcommand{\angstrom}{\text{\normalfont\AA}}

\title[Halo Assembly Bias of SDSS redMaPPer clusters]{Halo Assembly Bias using properties of central galaxies in SDSS redMaPPer clusters}

\author[T.~Sunayama et al.]{Tomomi Sunayama,$^{1}$\thanks{E-mail: sunayama.tomomi.f2@a.mail.nagoya-u.ac.jp}
Surhud More,$^{2,3}$
Hironao Miyatake,$^{1,2}$
\\
$^{1}$ Kobayashi-Maskawa Institute for the Origin of Particles and the Universe (KMI), Nagoya University, Nagoya, 464-8602, Japan\\
$^{2}$ Kavli Institute for the Physics and Mathematics of the Universe (WPI), The University of Tokyo Institutes for Advanced Study (UTIAS),\\The University of Tokyo, 5-1-5 Kashiwanoha, Kashiwa-shi, Chiba, 277-8583, Japan\\
$^{3}$The Inter-University Centre for Astronomy and Astrophysics, Post bag 4, Ganeshkhind, Pune 411007, India
}
\date{\today}
\pubyear{2022}

\begin{document}
\label{firstpage}
\pagerange{\pageref{firstpage}--\pageref{lastpage}}
\maketitle

\begin{abstract}
The clustering of dark matter halos depends on the assembly history of halos at fixed halo mass; a phenomenon referred to as \textit{halo assembly bias}. Halo assembly bias is readily observed in cosmological simulations of dark matter. However, it is difficult to detect it in observations. The identification of galaxy or cluster properties correlated with the formation time of the halo at fixed halo mass and the ability to select galaxy clusters free from projection effects are the two most significant hurdles in the observational detection of halo assembly bias. The latter, in particular, can cause a misleading detection of halo assembly bias by boosting the amplitude of lensing and clustering on large scales. This study uses twelve different properties of central galaxies of SDSS redMaPPer clusters derived from spectroscopy to divide the clusters into sub-samples. We test the dependence of the clustering amplitude on these properties at fixed richness. We first infer halo mass and bias using weak lensing signals around the clusters using shapes of galaxies from the SDSS survey. We validate the bias difference between the two subsamples using cluster-galaxy cross-correlations. This methodology allows us to decouple the contamination due to the projection effects from the halo assembly bias signals. We do not find any significant evidence of a difference in the clustering amplitudes correlated with any of our explored properties. Our results indicate that central galaxy properties may not correlate significantly with the halo assembly histories at fixed richness. %Among the twelve central galaxy properties that we test, $D_n 4000$ is the only parameter whose subsamples have equal mass but result in different clustering amplitude. The halo masses for the subsample split by $D_n 4000$ are ${\rm log}_{10} M_h=14.29 \msunh$, while the bias ratio $b_{\rm high}/b_{\rm low}=0.69 \pm 0.12$, an approximated 2$\sigma$ difference.
%However, we also find that the bias ratio measured from the clustering is $0.93 \pm 0.08$, and therefore we do not conclude that this is a detection of halo assembly bias. 
\end{abstract}

\begin{keywords}
large-scale structure of Universe -- galaxies: clusters: general -- cosmology: theory
\end{keywords}

% Enter the current year, for the copyright statements etc.

%%%%%%%%%%%%%%%%%%%%%%%%%%%%%%%%%%%%%%%%%%%%%%%%%%

%%%%%%%%%%%%%%%%% BODY OF PAPER %%%%%%%%%%%%%%%%%%

\section{Introduction}
In the cold dark matter (CDM) paradigm, massive cluster-sized halos form at rare peaks in the initial density field. Such massive halos are biased tracers of the matter density and cluster more strongly than the underlying matter distribution. The spatial distribution of dark matter halos primarily depends upon the halo mass \citep[e.g.,][]{Kaiser1984, Efstathiou1988, MoWhite1996}. However, there have been several studies that show that the clustering of halos at fixed halo mass can further depend upon secondary properties related to their assembly history \citep[e.g.,][]{gao_etal05, wechsler06, gao_white07, faltenbacher_white10, dalal_etal08}. This dependence of the halo clustering amplitude on secondary properties other than the halo mass is commonly referred to as \textit{halo assembly bias}, and can be traced back to the fact that halos of the same mass in different environments have different assembly histories and cluster differently. Having different assembly histories also affects the internal structure of halos \citep{bullock01,wechsler02,hahn_etal07,faltenbacher_white10}. This, in turn, results in a clustering dependence on the structural properties of a halo such as concentration. Even though halo assembly bias is sensitive to the specific definition of the property considered and can vary with halo mass \citep[e.g.,][]{Mao_etal2018,han_etal2019} , several studies reveal that the low-concentration halos, which are the late-forming halos, cluster more strongly than the high-concentration halos at massive end \citep[e.g.,][]{dalal_etal08}.

%{\bf These statements depend upon the definition of a formation time and the statements can be reversed in other definitions, so we should be careful.}

Since the halo assembly bias is a well-known phenomenon in simulations and has been studied extensively, several groups have attempted an observational detection of halo assembly bias with galaxies or galaxy groups, but they were inconclusive \citep[e.g.,][]{yang_etal06a, wang_etal13, Dvornik_etal2017, Niemiec_etal2018} or inconclusive \citep{Lin2016}.
The difficulty of detecting halo assembly bias arises from the search for observational proxies of the halo mass and the formation time of halos due to our inability to measure the structure and mass of individual halos directly. There are mainly two approaches used to define such proxies. One is to use the spatial distribution of member galaxies in groups and clusters, while another is to use the properties of central galaxies in groups/clusters. 

On galaxy cluster scales, \citet{Miyatake2016} and \citet{More2016} claimed evidence of halo assembly bias using the optically selected redMaPPer galaxy cluster catalogue \citep{Rykoff_etal2014}. 
They split the cluster sample based on the richness (i.e.,  the number counts weighted by their membership probabilities) as a proxy of halo mass and the compactness of member galaxies in the clusters as a proxy of halo concentration. 
They used weak gravitational lensing to confirm that the halo masses of these cluster subsamples were consistent with each other, and yet the clustering amplitudes were significantly different. The difference in 
the clustering amplitude of the two subsamples was almost $60\%$, much larger than the difference expected from N-body simulations. 
Because the detected signal was significantly larger than the signal expected from simulations, several studies investigated the cause of this significant signal \citep{Zu2017, BuschWhite2017, SunayamaMore2019}. 
All these studies concluded that the signal detected in \citet{Miyatake2016} and \citet{More2016} is likely a result of projection effects, where interloper galaxies along the line of sight to a cluster are mistaken as genuine members of the cluster.
The compactness of member galaxies, which was used as a proxy of concentration, was strongly correlated with the fraction of interloper galaxies. Therefore, using member galaxies in optically selected galaxy clusters to define the proxies of halo mass and formation time inevitably suffers from optical projection effects.

Using only central galaxies to detect the halo assembly bias is another way to avoid misidentifying member galaxies and the resulting contamination from the projection effects. \citet{Lin2016} used the galaxy group catalog from \citet{yang_etal06a}. They used the stellar mass and the specific star formation rate of central galaxies as proxies of halo mass and formation time, respectively. They split central galaxies into two subsamples such that they have equal distribution in stellar masses but different specific star formation rates. However, they attribute the difference in the clustering amplitude of the two subsamples to the halo mass difference of the subsamples using weak gravitational lensing measurements. Therefore, they do not claim evidence for the detection of halo assembly bias.

Recently, \citet{Zu_etal2021} investigated the origin of the scatter in the stellar-to-halo mass relation (SHMR) as well as the relation between concentration and stellar mass of the redMaPPer clusters, and found potential evidence of halo assembly bias. 
They first derived photometric stellar masses for all the brightest central galaxies (BCGs) in the redMaPPer cluster catalogue. They then split the sample into two subsamples based on their richness and stellar masses. 
The two subsamples were confirmed to have equal weak lensing halo masses but different concentrations and clustering amplitudes. 
The difference in the clustering amplitudes was about 10\%, albeit with a significant error, which is consistent with the size of cluster-sized halo assembly bias predicted from the simulations. 
Their result implies that the scatter in the stellar mass of the BCGs at fixed halo mass could be closely related to the halo concentration and the cosmic overdensities on large scales.

One caveat in this analysis is that the possible contamination from the projection effects, which biased the measurement of the clustering amplitude like \citet{Miyatake2016} and \citet{More2016} and is beyond the scope of their study.
The biased measurement due to the projection effects is a result of preferential selection of halos with filaments aligned along the line-of-sight (LOS) \citep{Sunayama_etal2020}. 
In other words, the projection effects cause anisotropies in the large-scale structure distribution around optically selected clusters and boost the amplitude of cluster lensing and clustering signals on large scales. 
Cluster-galaxy cross-correlations can be used to measure these anisotropies in the large-scale structure. 
However, \citet{Zu_etal2021} use photometric redshifts of BCGs to infer stellar masses.  21\% of the BCGs in their sample do not have spectroscopic redshifts and these BCGs without $z_{\rm spec}$ mostly have a stellar-mass below $10^{11}h^{-2}{\rm M_{\odot}}$ and some of them have above $3\times10^{11}h^{-2}{\rm M_{\odot}}$ \citep{Zu2020}. 
Including more low stellar mass BCGs implies more low-mass halos, whose richness could be affected by projection effects. 
In the study of \cite{Zu_etal2022}, they investigated why the stellar-mass split samples have almost equal halo masses but not the large scale biases and tested whether there is any evidence of projection effects in the large-scale bias measurement. They concluded that there is no strong contamination from the projection effects. 

The motivation of this study is to investigate further the possible correlation between central galaxy properties and the assembly history of halos in clusters while decoupling the physical correlation and the projection effects. To do so, we only use the redMaPPer clusters with spectroscopic redshifts. We will also discuss possible systematic biases due to spectroscopic incompleteness.

This paper is organized in the following manner. Section 2 describes the cluster and galaxy catalogues we use in our analysis. Section 3 presents the computation of cluster-galaxy cross-correlation functions and weak lensing measurements and the Bayesian inference methods to infer halo mass from the weak lensing measurements. We present our main results in Section 4 and conclude in Section 5.

\section{Data}
We use the redMaPPer cluster catalogue \citep{Rykoff_etal2014} derived from the Sloan Digital Sky Survey (SDSS) Data Release 8 (DR8) photometric galaxy catalogue \citep{Aihara_etal2011} and cross-match the central galaxies to the Wisconsin PCA-based value-added galaxy catalogue \citep{Chen_etal2012} to obtain galaxy properties for the central galaxies of the redMaPPer clusters. Also, we use the BOSS DR12 LOWZ sample to compute the cluster-galaxy cross-correlation functions.

\label{sec:data}
\subsection{redMaPPer galaxy clusters}

\begin{table}
	\centering
	\caption{List of central galaxy properties used to find the correlation with halo assembly history.}
	\label{tab:parameter}
	\begin{tabular}{cc} 
		\hline\hline
		Parameter & Description \\
		\hline
		${\rm D}_{n}4000$ & the 4000$\angstrom$ break strength\\
		$H{\delta}_{A}$ & the strength of Balmer absorption line\\
		$M_{*}/L$ & estimates of stellar mass-to-light ratio\\
		age & the r-band luminosity-weighted age\\
		$\mu \times \tau$ & the dust parameters $\mu$ and $\tau$\\
		$\sigma_v$ & the stellar velocity dispersion\\
		 $\Gamma_F$ &  fraction of stars formed in recent bursts \\
		 $Z$ & metallicity \\
		 $R_{{\rm deV},r}$ & de Vaucouleurs fit scale radius in $r$ band\\
		 $(a/b)_{{\rm deV},r}$ &  de Vaucouleurs fit a/b in $r$ band\\
		 $\phi_{{\rm deV},r}$ & de Vaucouleurs fit position angle in $r$ band\\
		$M_{*}$ & stellar mass\\
		\hline \hline
	\end{tabular}
	
\end{table}

The red-sequence Matched-filter Probabilistic Percolation (redMaPPer) cluster finding algorithm was developed by \citet{Rykoff_etal2014} to identify galaxy clusters using red-sequence galaxies in optical imaging surveys \citep{Rykoff_etal2014, Rozo2014, Rozo2015, Rozo2015_2}.
We use the redMaPPer galaxy cluster catalogue version 5.10 derived from the SDSS DR8 photometric galaxy catalogue \citep{Aihara_etal2011}.
The redMaPPer algorithm assigns redshift and richness, the number count of red galaxies weighted by their cluster membership probabilities, as well as centering probabilities to candidate central galaxies.
We choose galaxy clusters with richness $20\leq\lambda\leq200$ at $0.1 \leq z \leq 0.33$. This selection ensures that the redMaPPer cluster catalog is approximately volume-complete. The central galaxies are chosen based on the centering probabilities $p_{\rm cen}$ given by the redMaPPer algorithm. We select the galaxies with the highest $p_{\rm cen}$ as the central galaxies for our analysis. 
The reliability of the centering probability is studied by \cite{Hikage_2018}, which demonstrated that the strong consistency between the centring probabilities and clustering signals.
We limit ourselves to using the clusters whose central galaxies have a measured spectroscopic redshift and thus avoid uncertainties due to photometric redshift in the measurement of cluster lensing and clustering as well as to quantify the possible contamination by optical projection effects. 
In total, we have 7637 clusters which pass these cuts, while there are 1040 clusters without spectroscopic redshifts, merely $12\%$ of all the sample.
\footnote{In the original redMaPPer catalogue, there are 2009 central galaxies which do not have spectroscopic redshift. These galaxies correspond to  $23\%$ of all the clusters. To increase the number of the central galaxies with spectroscopic redshift, we carried out a further cross-match of the redMaPPer central galaxies to those in the SDSS DR17 catalogue. We thus obtain another 969 central galaxies with spectroscopic redshift.}

\begin{figure}
\centering
\includegraphics[width=0.45\textwidth]{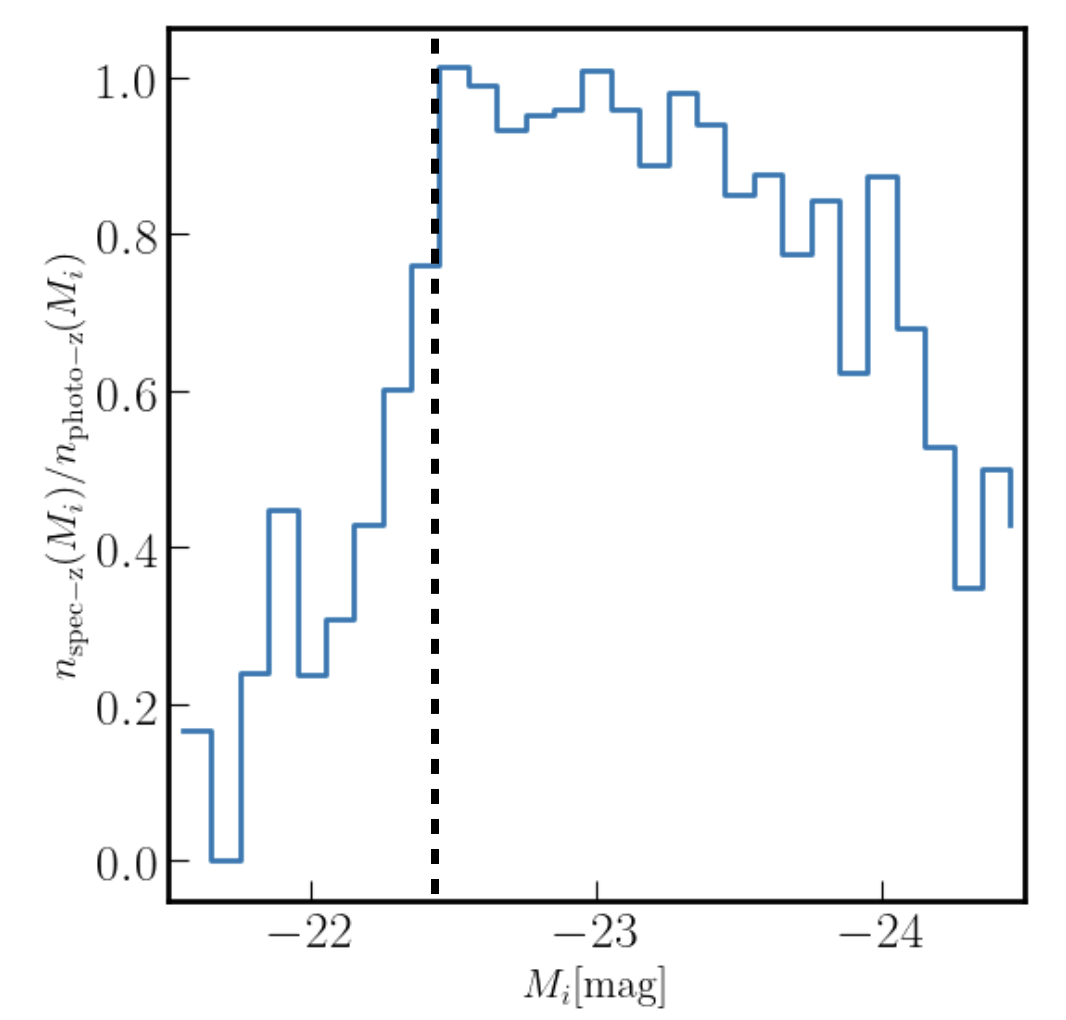}
\caption{\label{fig:mcut} We select redMaPPer clusters whose absolute magnitude is $M_{i}<-22.4$.}
\end{figure}

\cite{Zu2020} point out the spectroscopic incompleteness of the redMaPPer central galaxies -- the clusters without spectroscopic redshift are either the lowest or the highest stellar mass galaxies at fixed richness. To avoid such incompleteness, we decided to select further our cluster sample based on the absolute magnitude of the central galaxies. Fig.~\ref{fig:mcut} shows the detection fraction of the redMaPPer central galaxies by the spectroscopic observation as a function of $i$ band absolute magnitude $M_{i}$. As is consistent with Fig. 2 in \cite{Zu2020}, Fig.~\ref{fig:mcut} shows the declines towards zero at both ends of $M_{i}$. The cut-off at the high end of absolute magnitude is relatively sharp and happens at $M_{i}\sim -22.4$. To make sure that our cluster sample does not suffer from spectroscopic incompleteness, we only use the central galaxies whose absolute magnitude is $M_{i}\leq-22.4$. With this absolute magnitude cut, we lose another 4.5 per cent of the clusters, resulting in a sample of 7287 clusters we use in this study.
\footnote{We repeat the same analysis for the $M_*$-split subsamples without imposing the luminosity cut, and we did not find any significant changes in the result (see Appendix~\ref{app:no_lum}).}

To detect the halo assembly bias effect from the galaxy cluster sample, we need to split the sample into subsamples and find evidence for a dependence of the large scale clustering amplitude on any secondary property other than the halo mass. We match the most probable central galaxy of the redMaPPer clusters to the galaxies in the Wisconsin PCA-based value-added galaxy catalogue \citep{Chen_etal2012}. The value-added catalogue provides the values of each of the PCA coefficients, and we map these coefficients to 12 physical parameters based on the results of \citet{Chen_etal2012}. We list the corresponding physical parameters in Table~\ref{tab:parameter}, and we will use them as a proxy for concentration/formation time. This paper aims to find parameters that exhibit the assembly bias effect. We use ten equally spaced bins both in redshift and richness to obtain the median value of the chosen parameter X. We use a spline fit for the median of $X$ as a function of redshift and richness. Based on this fit, we define the two subsamples as "large-$X$ clusters" with $X>\overline{X}(z,\lambda)$ and "small-$X$ clusters" with $X<\overline{X}(z,\lambda)$. This selection makes sure that the two cluster subsamples have the same richness and redshift distributions. 

We use the random catalogues corresponding to the redMaPPer cluster catalogue for the clustering and weak lensing measurements. These catalogues contain corresponding position information, redshift, richness, and a weight for each random cluster.

\subsection{BOSS DR12 LOWZ sample}
We will cross-correlation of the redMaPPer galaxy clusters with spectroscopic galaxies to study halo assembly bias. 
We use the spectroscopic galaxies in the large-scale structure catalogues constructed from SDSS DR12 (\citet{Alam2015}). In particular, we will use the LOWZ sample since it has considerable overlap in the redshift range with our galaxy cluster sample. We restrict ourselves to LOWZ galaxies with redshifts between $0.1<z<0.33$, the same redshift range as our galaxy clusters.  
The galaxy catalogue also comes with associated random galaxy catalogues that we use to perform our cross-correlation analysis.

\section{Methods}

\subsection{Detecting Halo Assembly Bias Without Projection Effects}
\label{sec:method:basic}

\begin{figure*}
\centering
\includegraphics[width=0.95\textwidth]{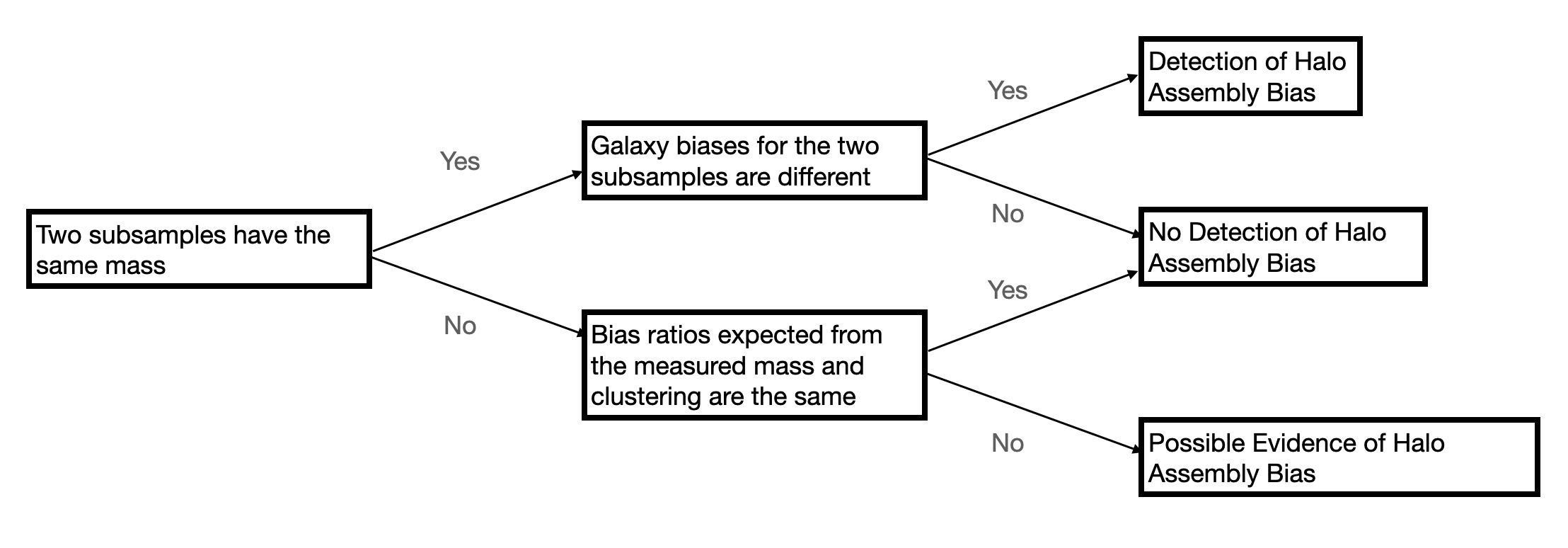}
\caption{\label{fig:logic1} The logic to detect halo assembly bias often used in other studies.}
\end{figure*}

\begin{figure*}
\centering
\includegraphics[width=0.95\textwidth]{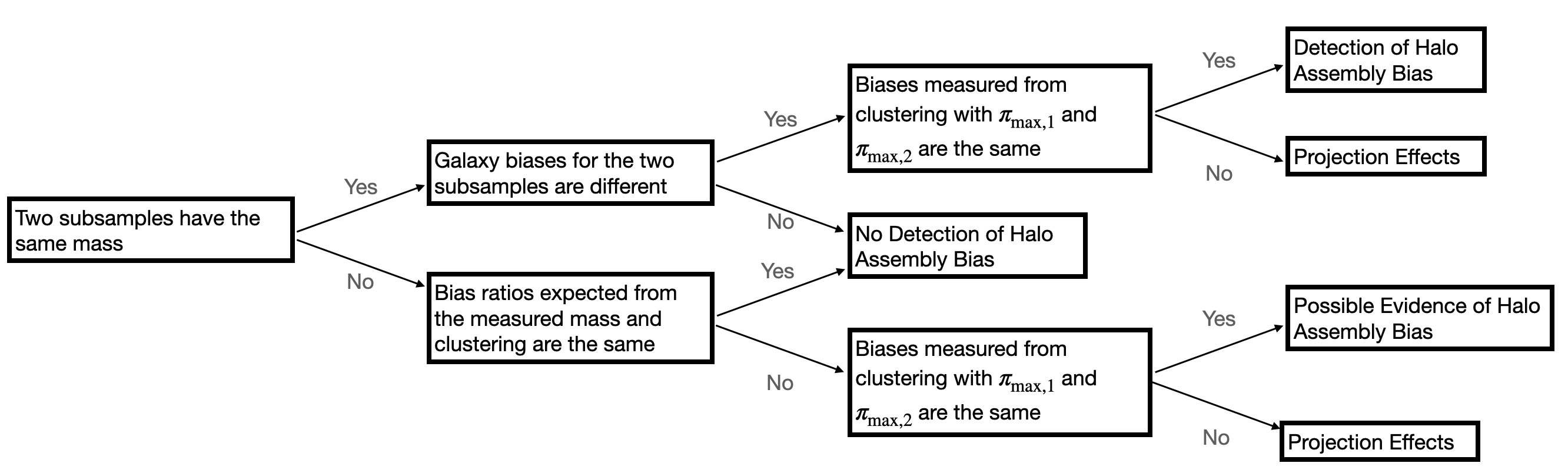}
\caption{\label{fig:logic2} The logic to detect halo assembly bias by decoupling from the projection effects.}
\end{figure*}

We take a two-step approach to detect halo assembly bias using cluster subsamples - i) we confirm that the difference in the biases is not due to halo mass differences in the subdivided samples, or, ii) arising due to contamination of the cluster subsamples due to LOS projection effects. We use the weak gravitational lensing signal to measure the halo mass for two cluster subsamples split based on the parameter $X$. Although the weak lensing signal can also be used to infer the large scale bias of the cluster subsamples, the weak lensing signal (or the cluster-cluster auto-correlation) has a lower signal-to-noise ratio than the cross-correlation of the cluster subsamples with galaxies. Therefore we use the galaxy cluster-galaxy cross-correlation to detect the halo biases of the two sub-samples.

Fig.~\ref{fig:logic1} shows the simple logical flow used in previous studies to detect halo assembly bias. If the two subsamples have equal halo masses but have different halo biases, it is typically considered evidence favouring halo assembly bias. However, this test on its own does not guarantee that the observed bias difference between the two sub-samples could be due to projection effects \citep[e.g.,][]{Miyatake2016}.
To decouple the bias difference caused by the projection effects from the halo assembly bias, we instead use the logical flow described in Fig.~\ref{fig:logic2}. The first step in this revised logical flow is the same as that shown in Fig.~\ref{fig:logic1}: measure the halo masses of the two subsamples from lensing and the bias from cross-correlations. 

However, our new logical flow includes an extra check. We compute the cluster-galaxy projected cross-correlations with two different integration limits $\pi_{\rm max,1}$ and $\pi_{\rm max,2}$. The projected cross-correlation function is given by,
\begin{equation}
w_{\rm p,cg}(R, \pi_{\rm max}) = 2\int_{0}^{\pi_{\rm max}}\xi_{\rm cg}(R, \pi) d\pi\,.
\end{equation}
In the absence of any projection effects, or redshift space distortions, the quantity $\xi_{\rm cg}(R, \pi)$ in the above equation would be equal to the 3-dimensional cross-correlation $\xi_{\rm cg}\left[(R^2+\pi^2)^{1/2}\right] \propto b_{\rm c}b_{\rm g}$. The ratio of the cross-correlations of galaxies with the two cluster subsamples will be equal to $b_{\rm c1}/b_{\rm c2}$.

Redshift space distortions ought to cancel out in the limit that $\pi_{\rm max}\rightarrow \infty$. However as pointed out in \citet{vdBosch_etal2013}, the finite integration length can result in residual redshift space distortions in a scale dependent manner. For $R/\pi_{\rm max}\leq 1$, the residual RSDs are small, while these corrections become larger on scales $R/\pi_{\rm max}\sim 1$. By using the formalism in \citet{vdBosch_etal2013}, and using fiducial bias ratios $3$ and $2.5$, we have computed the effect of the residual redshift space distortions on the ratio of the biases for $\pi_{\rm max}=[50, 100] \mpch$. Such residual redshift space distortions can cause a scaled dependent bias ratio but within $\sim 2$ percent, which is much smaller than our errors.

Projection effects in the sample selection can cause an anisotropic distribution of matter/galaxies around clusters \citep{BuschWhite2017, SunayamaMore2019, Sunayama_etal2020, Park_etal2021}. Such projection effects can lead to differences between the structure of the correlation function in redshift space. This anisotropic distribution lead to projected cluster-galaxy cross-correlations that depend upon the LOS integration length scale \citep{Sunayama_etal2020}. Differences in projection effects can amplify the inferred bias ratios of the cluster subsamples, when smaller projection lengths ($\pi_{\rm max}=50 \mpch$ versus $100\mpch$) are used compared to larger ones.

Ignoring residual redshift space distortions, in the presence of projection effects,
\begin{align}
w_{\rm p,cg}(R,\pi_{\rm max})=[1+\Pi(\pi_{\rm max})]w_{\rm iso,cg}(R)
\label{eq:inertia}
\end{align}
where $\Pi$ is the anisotropic enhancement of the clustering amplitude,
%and $\pi_{\rm max}$ is the integral scale used to compute the projected correlation function from 3D correlation function $\xi(r)$:
%\begin{equation}
%w_{\rm p}(R) = 2\int_{R}^{\infty}\xi(r)\frac{r %dr}{\sqrt{(r^2-R^2)}}.
%\label{eq:abel}
%\end{equation}
and $w_{\rm iso, cg}(R)$ is the expected projected cross-correlation function of galaxies and clusters, in the absence of any projection effects that correlated with the property used for subsample selection. This $\Pi(\pi_{\rm max})$ can be modeled as constantly increasing function of $\pi_{\rm max}$ \citep{Park_etal2021}.
If one cluster subsample is heavily affected by the projection effects (denote as "sample1") but not the other (denote as "sample2"), then the bias ratio between these two subsamples will show a strong dependence on the choice of the integral scale such as 
\begin{align}
\frac{b_{\rm sample1}}{b_{\rm sample2}}(\pi_{\rm max})=(1+\Pi_{\rm sample1}(\pi_{\rm max}))\frac{b_{\rm sample1,iso}}{b_{\rm sample2,iso}},
\label{eq:inertia}
\end{align}
where $b_{\rm sample,iso}$ is an expected bias for the case of isotropically distributed clusters with the same halo masses and $b_{\rm sample2}=b_{\rm sample2,iso}$.
If two cluster subsamples are equally affected by the projection effects, the effects will be canceled out such as
\begin{align}
\frac{b_{\rm sample1}}{b_{\rm sample2}}(\pi_{\rm max})=\frac{(1+\Pi_{\rm sample1}(\pi_{\rm max}))}{(1+\Pi_{\rm sample2}(\pi_{\rm max}))}\frac{b_{\rm sample1,iso}}{b_{\rm sample2,iso}}=\frac{b_{\rm sample1,iso}}{b_{\rm sample2,iso}},
\label{eq:inertia}
\end{align}
if $\Pi_{\rm sample1}(\pi_{\rm max})=\Pi_{\rm sample2}(\pi_{\rm max})$.

Therefore, we study the dependence of the bias ratio. We will use any statistically significant dependence of the bias difference on the choice of the integral scale $\pi_{\rm max}$ to flag projection effects. If such a dependence is not found, a difference in the halo bias between two subsamples at fixed weak lensing halo mass can be construed as evidence of halo assembly bias.
%Projection effects cause an anisotropic boost on the amplitude of clustering and hence show a dependence on the choice of the integration scale \citep{Sunayama_etal2020}.

\subsection{Galaxy-Galaxy Lensing Measurements}
\label{sec:lensing}
Weak gravitational lensing provides a way to infer the mean halo masses of galaxy clusters and bias on large scales.
The excess surface mass density profile $\Delta\!\Sigma$, which is a direct observable of cluster lensing, is given by
\begin{align}
\Delta\! \Sigma(R)&\equiv \avrg{\Sigma}\!(<R)-\Sigma(R)\nonumber\\
&= \frac{2}{R^2} \int_0^R\! \mathrm{d}R' R' \Sigma(R') - \Sigma(R), 
\label{eq:DeltaSigma}
\end{align}
where $\avrg{\Sigma_i}\!(<R)$ and $\Sigma_{i}(R)$ denote the average surface mass density within a circle of radius $R$ and the surface mass density at $R$ respectively. 
The surface mass density $\Sigma_{i}(R)$ is computed from the halo-mass cross correlation function $\xi_{\rm cm}(r)$,

\begin{align}
\Sigma(R)=\bar{\rho}_{\rm m0}\int_{-\infty}^{\infty}\!\!\mathrm{d}\pi~
\left[1+\xi_{\rm cm}\!\left(\sqrt{R^2+\pi^2}\right)\right],
\label{eq:Sigma}
\end{align}
where $\bar{\rho}_{\rm m0}$ is the mean matter density today, $\pi$ and $R$ are the separations parallel and perpendicular to the LOS direction from the cluster center, respectively.

%The weak gravitational lensing signal at the projected separation from the cluster center R is given by
%\begin{equation}
%\Delta \Sigma(R) = \bar{\Sigma}(<R)-\Simga(R)=\Sigma_{\rm crit} \gamma_{\rm t},
%\end{equation}
%where $\Delta \Sigma(R)$ is the excess surface density, $\Sigma(R)$ is the projected mass density profile, $\bar{\Sigma}(<R)$ is the average mass density within R, $\Sigma_{\rm crit}$ is the critical surface mass density, and $\gamma_{\rm t}$ is the tangential shear. 

%The critical density $\Sigma_{\rm crit}$ for lens and source galaxies depends on the redshifts $z_l$ and $z_s$ respectively and is given by
%\begin{equation}
%\Sigma_{\rm crit} = \frac{c^2}{4\pi G}\frac{d_A(z_s)}{(1+z_l)^2 d_A(z_l)d_A(z_l,z_s)},
%\end{equation}
%where $d_A(z_l)$, $d_A(z_s)$, and $d_A(z_l,z_s)$ are the comoving angular diameter distances for the lens at $z_l$, the source at $z_s$,and the lens-source pair from $z_l$ to $z_s$, respectively. The factor $(1+z_l)^2$ in the numerator arises from the choice of comoving coordinates.

To compute the weak lensing signal for our cluster sample, we follow the methodology described in \citet{Mandelbaum_etal2013} and \citet{Miyatake2016}. We use the shape catalogue of \citet{Reyes_etal2012} which is based on the photometric galaxy catalogue from the SDSS DR8. The shape measurements were calibrated using detailed image simulations \citep[see][for details]{mandelbaum_etal2012}. The galaxy shapes are measured by the re-Gaussianization technique \citep{hirata_seljak03}. \citet{mandelbaum05} extensively investigate the systematic uncertainties in the shape measurements of SDSS galaxies. The redshift of source galaxies is estimated based on the photo-$z$ code ZEBRA \citep{Feldmann2006, Nakajima_etal2012}. 

We measure the average projected mass density profile $\Delta \Sigma(R)$ as
\begin{equation}
\Delta \Sigma(R)=\frac{\Sigma_{ls} w_{ls} e_{t,ls}\left<\Sigma_{\rm crit}^{-1}\right>_{ls}^{-1}}{(1+\hat{m}(R))\Sigma_{ls} w_{ls}},
\end{equation}
where the summation runs over all the lens-source pairs separated by R and the subscript "$l$" and "$s$" represent lens and source galaxies, respectively. $e_{t,ls}$ is tangential components of ellipticities and $w_{ls}$ is given by
\begin{equation}
w_{ls} = w_l w_s \left<\Sigma_{\rm crit}^{-1}\right>_{ls}^{2},
\end{equation}
where $w_l$ and $w_s$ are the weight given by the redMaPPer cluster catalog and the source catalog, respectively. The overall factor $(1+\hat{m}(R))$ is to correct for a multiplicative shear bias \citep{miller12} and is computed as an average of the lens-source pairs at a distance $R$.

%photo-z correction
To measure the lensing signal accurately, we apply several corrections to remove the residual systematics.
The first of such correction corresponds to the photometric redshift bias following Eqn. 5 in \cite{mandelbaum08a}. When $z_l$ and $z_s$ are photometric redshifts, the measured critical surface density $\Sigma_{\rm crit}$ is biased such as
\begin{equation}
1+b_z(z_l) =\frac{\Delta \Sigma}{\tilde{\Delta \Sigma}}=\frac{\Sigma_{s} w_{ls} \Sigma_{\rm{crit},ls} \tilde{\Sigma}_{\rm{crit},ls}^{-1}}{\Sigma_{s} w_{ls}},
\end{equation}
where $\tilde{\Sigma}_{\rm{crit},ls}$ is the true critical surface mass density measured by \cite{Nakajima_etal2012} for which we used spectroscopic redshifts for source galaxies. This quantity $1+b_z$ is a function of lens redshift $z_l$. We measured the photo-$z$ bias as a function of lens redshift and computed an average photo-$z$ bias for the redMaPPer clusters. The average value of the bias parameter over the redshift range of our lensing sample is about $8.3\%$. We divide the measured lensing signal by $1+b_z$.

%boost factor
The second is a dilution of the lensing signal due to source galaxies which are physically associated with the lens galaxy and can be corrected by a boost factor $B(R)$ given by
\begin{equation}
B(R)=\frac{\sum_i w_r}{\sum_i w_l} \frac{\Sigma_{ls} w_l w_s}{\Sigma_{rs} w_r w_s},
\end{equation}
where the superscript “$r$” stands for random catalogs and $w_r$ is the weight of the random provided by the catalog.
If there is no physical correlation between lens and source galaxies, $B(R)=1$. The boost factor $B(R)$ deviates from 1 on small scales but converges to 1 on large scales. 
By multiplying the measured signal by the boost factor, it corrects for the dilution of the signal.

%random signal
The third is a possible residual systematic in the shape measurements due to imperfect optical distortion corrections across the field of view. If we measure the lensing signal around the random points, we should not detect any coherent tangential distortions. Any coherent tangential distortion around the randoms arises due to imperfect corrections that translate into systematics in the shape measurements. We correct for such effects by subtracting the lensing signal around the randoms from that around the lenses.

We perform halo model fits to the measurements of the weak lensing signal of each cluster subsample. 
To model $\Delta\! \Sigma(R)$, we use a halo model assuming NFW profile with the off-centering effect. 
We use a five parameter model to the weak lensing signal following \cite{Miyatake2016},
\begin{align}
\Delta\! \Sigma(R; M_{\rm 200m}, c_{\rm 200m}, q_{\rm cen},\alpha_{\rm off},b)\nonumber\\
= q_{\rm cen} \Delta\! \Sigma^{\rm NFW}(R;M_{\rm 200m},c_{\rm 200m},\alpha_{\rm off})\nonumber\\
+(1-q_{\rm cen})\Delta\! \Sigma^{\rm NFW,off}(R;M_{\rm 200m},c_{\rm 200m},\alpha_{\rm off}) \nonumber\\
+\Delta\! \Sigma^{*}(R;M_{*})+\Delta\! \Sigma^{\rm 2-halo}(R;b). 
\label{eq:dSigma_model}
\end{align}
The first term is the halo mass profile for the clusters whose centre is correctly identified, while the second term is the halo mass profile for these clusters whose centres are misidentified.
The parameter $1-q_{\rm cen}$ represents the fraction of the off-centring clusters, and $\alpha_{\rm off}$ describes the ratio of the three-dimensional off-centring radius to $R_{\rm 200m}$.
The normalized profile of the off-centering clusters with respect to their true center is given by $u_{\rm off}(r) \propto \exp[-r^2/(2\alpha_{\rm off}^2 R_{\rm 200m}^2)]$.
Note that we truncate the off-centring profile to zero at $r>R_{\rm 200m}$.
The third term models a stellar-mass contribution from the central galaxies assuming a point mass, and we use the mean value of $M_{*}$ in the cluster subsamples.
The fourth term models the lensing contribution from two-point correlation functions between the clusters and the mass distribution around the clusters, which is given by
\begin{align}
\Delta\! \Sigma^{\rm 2-halo}(R)=b \int k dk/(2\pi) \bar{\rho}_{\rm m0} P_{\rm m}^{\rm L}(k,z_{\rm cl})J_2(kR),
\label{eq:two-halo}
\end{align}
where $b$ is a linear galaxy bias, $P_{\rm m}^{\rm L}$ is the linear matter power spectrum and $J_2(x)$ is the second-order Bessel function.

%We perform Bayesian parameter inferences with the likelihoods {\bf likelihood needs to be defined.} to obtain the posterior distribution of the parameters. 
%To do that, we use the affine invariant Monte Carlo Markov Chain (MCMC) ensemble sampler \textit{emcee} \citep{emcee}.
We perform Bayesian parameter inferences using the likelihoods with the measured lensing signal and the model prediction as in Eqn.~\ref{eq:dSigma_model} to obtain the posterior distribution of the parameters. 
To do that, we use the affine invariant Monte Carlo Markov Chain (MCMC) ensemble sampler \textit{emcee} \citep{emcee}.
We use flat priors for all the parameters: $M_{\rm 200m}/[]10^{14}\msunh \in [0.5,50]$, $c_{\rm 200m} \in [1,10]$, $q_{\rm cen} \in [0.6,1]$, $\alpha_{\rm off} \in [10^{-4},1]$, and $b \in [0,10]$.

%We use flat priors for halo mass $M$ and bias $b$ and Gaussian priors for concentration $c$, $q_{\rm cen}$, and $\alpha_{\rm off}$. 
%This is to break the degeneracy between halo concentration $c$ and off-centering parameters $1-q_{\rm cen}$ and $\alpha_{\rm off}$. Based on the X-ray observations from \textit{Chandra} \citep{Zhang_etal2019}, the average off-centering fraction are $\left <1-q_{\rm cen} \right >=0.3 \pm 0.04$ and $\left <\alpha_{\rm off} \right >=0.18 \pm 0.02$. 
%Since these average off-centering parameter values are for the entire cluster sample, we jointly fit the lensing profiles for the two subsamples following \cite{Zu_etal2021}. {\bf Tomomi, the model used by Zhang is quite different from the model assumed in the 3d gausssian case here. Are you sure we can use these priors?}
%We use $c \sim \mathcal{N}(5,1.5^2)$, $1-q \sim \mathcal{N}(0.3,0.04^2)$, and $\alpha_{\rm off} \sim \mathcal{N}(0.18,0.02^2)$.
%\footnote{We repeat the same analysis with the flat priors for the off-centering parameters and the concentration in Appendix~\ref{app:flat}.}

The background cosmology assumed for the lensing and clustering measurements is based on \textit{Planck} 2018 \cite{Planck2018} under a flat $\Lambda$CDM model: $\Omega_{\rm m}=0.315$ and $\sigma_8=0.811$.

\subsection{Correlation function measurements}
\label{sec:corr}
Once we find the cluster subsamples with equal halo masses but different large scale amplitudes, we examine whether the clustering difference indicates halo assembly bias or is a result of projection effects. We use the cluster-galaxy cross-correlation function to differentiate between these possibilities.
The cluster-galaxy cross-correlation function can be computed using the Landy-Szalay estimator \citep{landyszalay93} as,
\begin{equation}
\xi_{\rm cg}=\frac{\left <D_c D_g  \right >-\left <D_g R_c \right >-\left <D_c R_g \right >+\left <R_c R_g \right >}{\left <R_c R_g \right >},\label{eq:landy-szalay}
\end{equation}
where $\left <DD \right >$,$\left <DR \right >$, and $\left <RR \right >$ are the normalized numbers of pairs for data-data, data-random, and random-random respectively.
The subscripts $c$ and $g$ represent the samples of clusters and galaxies.
We measure the cluster-galaxy cross-correlation functions $\xi_{cg}$ in two dimensions $(R,\pi)$ where $R$ and $\pi$ are the separations perpendicular to and along the LOS direction.

Then, we compute the projected correlation functions by integrating $\xi_{cg}$ along the LOS direction up to the maximum integral scale $\pi_{\rm max}$,
\begin{equation}
w_{\rm p}(R,\pi_{\rm max})=\int_{0}^{\pi_{\rm max}}d\pi \xi_{\rm cg}(R,\pi)\,.\label{eq:wp_cg}
\end{equation}
We use $\pi_{\rm max}=50h^{-1}{\rm Mpc}$ and $100h^{-1}{\rm Mpc}$.
If the bias difference between two subsamples is due to the projection effects, the bias ratios $w^{\rm large}_{\rm p,cg}(R)/w^{\rm small}_{\rm p,cg}(R)$ will have different values for a different choice of $\pi_{\rm max}$. 
To check this, we compute the ratio of ratio $R=\frac{w^{\rm large}_{\rm p,cg1}}{w^{\rm small}_{\rm p,cg1}}/\frac{w^{\rm large}_{\rm p,cg2}}{w^{\rm small}_{\rm p,cg2}}$, where the subscripts $1$ and $2$ correspond to the used maximum integral scales $\pi_{\rm max}=50h^{-1}{\rm Mpc}$ and $100h^{-1}{\rm Mpc}$.
If this $R$ is unity, the bias difference between the two subsamples is not likely due to projection effects and therefore signifies evidence for halo assembly bias.

\section{Results}
\label{sec:results}

In this section, we first present our approach to the detection of halo assembly bias for the case of stellar mass $M_*$ in Sec.~\ref{sec:res:mstar} and then we show the results for other parameters listed in Table~\ref{tab:parameter} in Sec.~\ref{sec:res:others}.

\subsection{Subsamples divide by BCG stellar mass}
\label{sec:res:mstar}

\begin{figure*}
\centering
\includegraphics[width=0.8\textwidth]{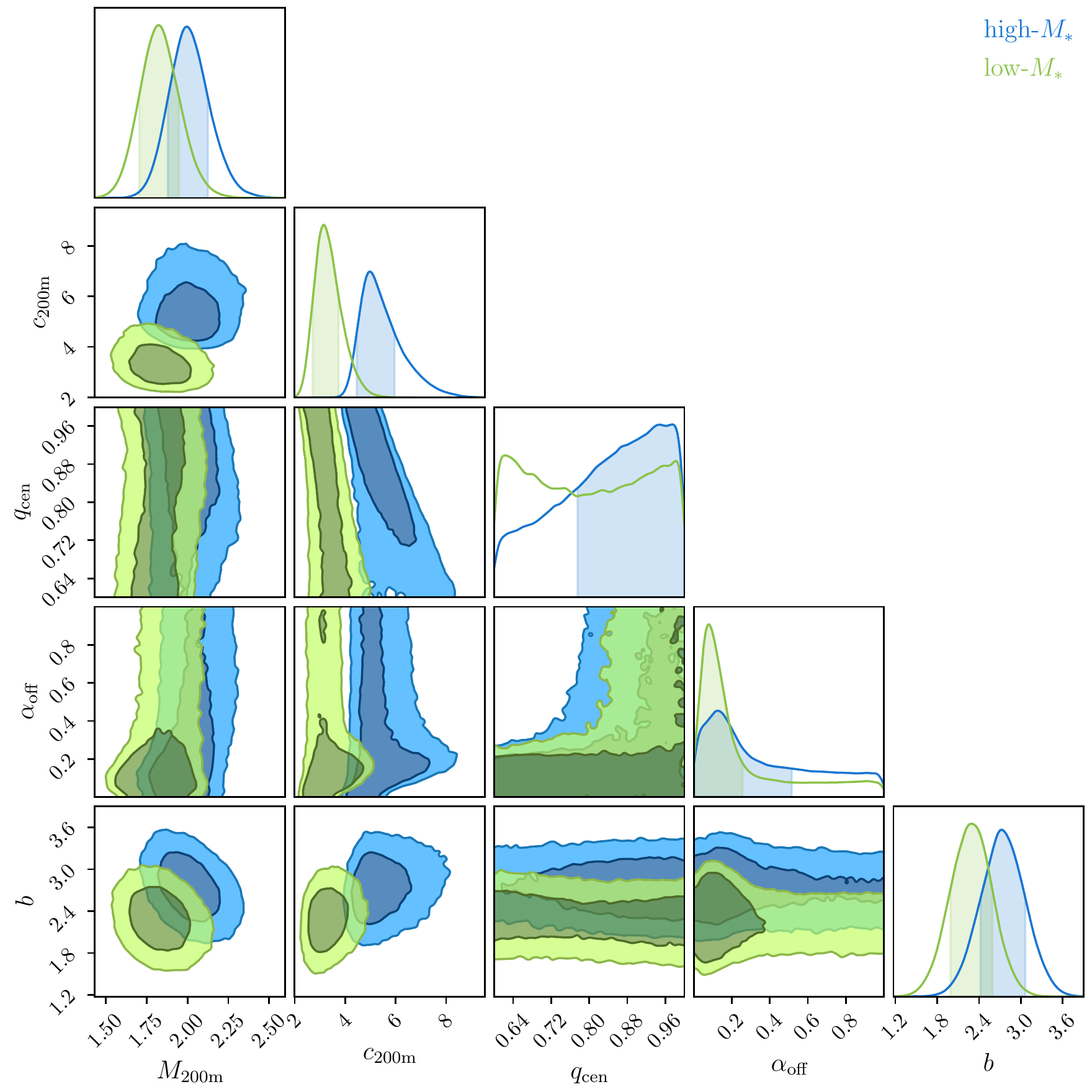}
\caption{\label{fig:triangle_mstar}  Posterior constraint from the modelling of $\Delta \Sigma(R)$ for the high-$M_*$ (blue) and low-$M_*$ (green) subsamples. Each histogram in the diagonal panels are the 1D marginalised posterior distribution of the parameters, while each contour plot in the off-diagonal panels indicate the 1$\sigma$ and 2$\sigma$ confidence regions of the matching parameter pairs.}
\end{figure*}

We show the posterior distribution of our model parameters inferred from the measurement of the cluster subsamples divided by high-$M_*$ (blue) and low-$M_*$ (green) of the BCGs in Fig.~\ref{fig:triangle_mstar}. 
The diagonal panels show the marginalized posterior distributions for the five parameters used in the model, while the off-diagonal panels are the 1$\sigma$ and 2$\sigma$ credible regions for the parameter combinations. The mean halo masses of the high-$M_*$ and low-$M_*$ subsamples are ${\rm log}_{10} M_h^{\rm high}=14.30 \pm 0.03 \msunh$ and ${\rm log}_{10} M_h^{\rm low}=14.28 \pm 0.03 \msunh$, respectively. At fixed $\lambda$, the result implies that the two stellar mass BCG subsamples reside in halos of similar masses at fixed richness \citep{Zu_etal2021, Zu_etal2022}.

The measured halo bias ratio from lensing signals is $b_{\rm high}/b_{\rm low}=1.20\pm 0.12$, while the expected halo bias ratio based on the measured halo masses using \textit{Colossus} is $b_{\rm high}/b_{\rm low}= 1.04 \pm 0.03$.%{\bf Please add an errorbar on this expectation by propagating the error on halo masses using the MCMC.}
The measured bias ratio is consistent with the expected bias ratio based on the estimate of the halo masses, implying that this does not constitute substantial evidence of halo assembly bias.
On careful inspection, this seems consistent with the results shown in Fig.~4 of \cite{Zu_etal2021}; within the errors, the mass and bias differences between the two subsamples are consistent. 
%Our measured halo masses and halo biases for the high/low-$M_*$ subsamples indicate that the stellar mass is correlated with halo mass.
\cite{Zu_etal2021} further show that the cross-correlation between SDSS photometric galaxies and the cluster subsamples shows a difference in clustering of order $10\%$. On the other hand, the cross-correlation with spectroscopic galaxies gives a noisy result. Given the larger errors, we cannot entirely rule out such a difference.

We do not, however, see consistent differences in the inferred posteriors of concentration and bias (contrast the second panel in the bottom row of Fig.~\ref{fig:triangle_mstar} with the bias-concentration subpanel in Fig. 4 of \citet{Zu_etal2021}). One of the differences between our sample selection and that of \cite{Zu_etal2021} is that they use all redMaPPer clusters with $\lambda \geq 20$ at $0.17 <z< 0.3$, while we limit ourselves to the clusters with spectroscopic redshift and with $M_i \leq -22.4$. As we would like to test for projection effects, we do not conduct a cross-correlation analysis with photometric galaxies where we cannot test for projections.
\footnote{We tested a similar cluster sample selection as is \cite{Zu_etal2021}. We select redMaPPer clusters at $0.17 \leq z_{\lambda} \leq 0.3$ with richness $\lambda \geq 20$ and the luminosity cut of $M_i \leq -22.4$ and repeat the analysis. We obtained a similar result that low-$M_*$ subsample has a lower value of concentration and with similar halo masses and halo biases.}
%It means that the cluster sample used in the analysis of \cite{Zu_etal2021} contains more low mass halos than our sample, and it can be a reason of the difference.

%The measured biases are significantly lower than the expected values, however, these results are consistent with the results in \cite{Zu_etal2021} where the measured halo bias from lensing signals is smaller than the prediction.
%Although our measured halo biases are consistent with the values from \cite{Zu_etal2021}, our selection of the redMaPPer clusters and the results of the halo masses are very different from \cite{Zu_etal2021}.
%We limit our cluster sample to be spectroscopically observed and only select the ones with $M_{i}<-22.4$, while \cite{Zu_etal2021} used all the redMaPPer clusters with $\lambda>20$ at $0.17 \leq z \leq 0.3$.
%We see a positive correlation between halo mass and stellar mass, but \cite{Zu_etal2021} did not find such correlation.
%Since \cite{Zu_etal2021} uses all the redMaPPer clusters with $\lambda>20$, one of the subsamples may contain more low-mass halos than the other due to the projection effects, and these low-mass halos can possibly lower the mean halo mass of that subsample.

While the fractions of centered clusters $q_{\rm cen}$ are consistent with the fraction of the clusters with the centering probability $p_{\rm cen} \geq 0.8$, which is roughly $\sim 76\%$, our results show a degeneracy between the concentration parameter $c$ and the off-centering kernel size $\alpha_{\rm off}$. The degeneracy results in $c^{\rm high}=5.31 \pm 1.01$ and $c^{\rm low}=3.28 \pm 0.60$. The expected values of concentration using the model by \cite{diemer_joyce2019} from \textit{Colossus} are $c^{\rm high}=4.94 \pm 0.02$ and $c^{\rm low}=4.98 \pm 0.03$.
Since the low-$M_*$ subsample contains more low-mass clusters, $c^{\rm low}$ is expected to be slightly larger than $c^{\rm high}$.
The cause of this opposite result can be a degeneracy with the off-centring parameter $\alpha_{\rm off}$.
To suppress the contribution from the off-centred clusters, we select the redMaPPer clusters whose central probability $p_{\rm cen}$ is larger than 0.9 and repeat the same analysis.
\footnote{Posterior constraints of the model parameters for the cluster subsamples with $p_{\rm cen}\geq0.9$ are in Appendix~\ref{app:pcen}.}
While we do see the increase of $q_{\rm cen}$ to $0.95^{+0.04}_{-0.03}$ for both subsamples (i.e., decrease in the fraction of the off-centered clusters) and the shift of the off-centering location $\alpha^{\rm high}_{\rm off}=0.55^{+0.31}_{-0.35}$ and $\alpha^{\rm low}_{\rm off}=0.51^{+0.34}_{-0.34}$, we do not see significant changes in concentration, $c_{\rm high}=4.91^{+0.50}_{-0.45}$ and $c_{\rm low}=3.57^{+0.45}_{-0.41}$.

\begin{figure}
\centering
\includegraphics[width=0.45\textwidth]{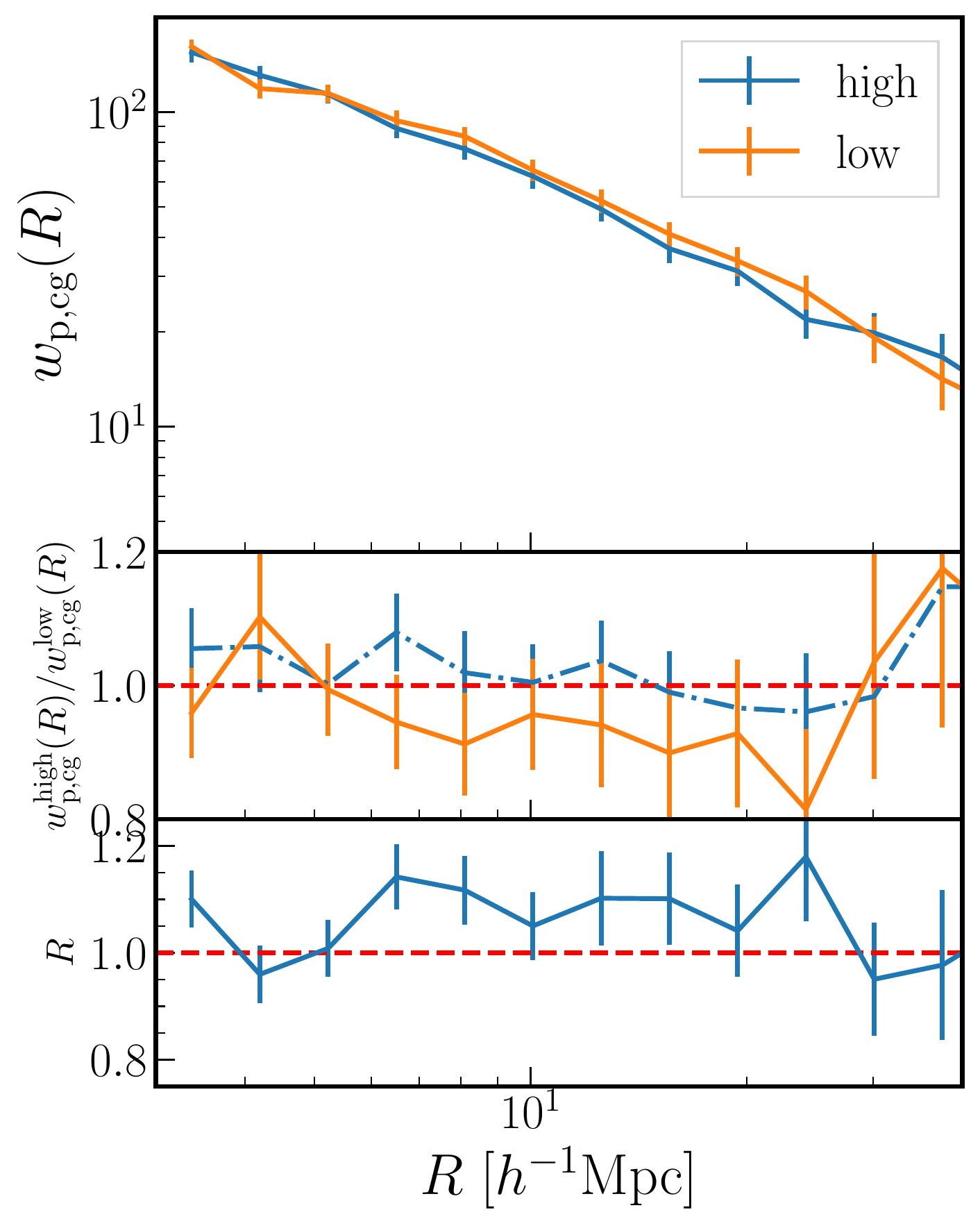}
\caption{\label{fig:ror_mstar} Top: Cluster-galaxy projected cross-correlation functions integrated up to $\pi_{\rm max}=100h^{-1}{\rm Mpc}$ for the high/low-$M_*$ subsamples. Middle:The ratio $w_{\rm p,cg}^{\rm high}(R)$ and $w_{\rm p,cg}^{\rm low}(R)$ computed with $\pi_{\rm max}=50h^{-1}{\rm Mpc}$ (blue dash-dot) and $\pi_{\rm max}=100h^{-1}{\rm Mpc}$ (orange solid). Bottom: The ratio of $w_{\rm p,cg}^{\rm high}(R)/w_{\rm p,cg}^{\rm low}(R)$ at $\pi_{\rm max}=50h^{-1}{\rm Mpc}$ and $\pi_{\rm max}=100h^{-1}{\rm Mpc}$ denoted as $R$ which is defined in Sec.~\ref{sec:corr}.}
\end{figure}

To examine whether the $M_*$-split subsamples are affected unevenly by projection effects, we compute the cluster-galaxy projected cross-correlation functions $w_{\rm p,cg}(R)$ with different LOS integral lengths $\pi_{\rm max}=50h^{-1}{\rm Mpc}$ and $100h^{-1}{\rm Mpc}$, respectively. 
The bottom panel of Fig.~\ref{fig:ror_mstar} shows the ratio of $w_{\rm p,cg}^{\rm high}(R)/w_{\rm p,cg}^{\rm low}(R)$ at $\pi_{\rm max}=50h^{-1}{\rm Mpc}$ and $100h^{-1}{\rm Mpc}$ denoted as $R$.
If this ratio is one, the bias difference between the subsamples is less likely to result from projection effects.
However, Fig.~\ref{fig:ror_mstar} shows $1\sigma$ deviation from unity, which is $1.09 \pm 0.09$ on $10\mpch \leq R \leq 30 \mpch$. 
This result implies that the high/low-$M_*$ subsamples are likely affected differently by projection effects.
The middle panel of Fig.~\ref{fig:ror_mstar} shows the ratio $w_{\rm p,cg}^{\rm high}(R)/w_{\rm p,cg}^{\rm low}(R)$ with $\pi_{\rm max}=50h^{-1}{\rm Mpc}$ (blue dashed-dot) and $\pi_{\rm max}=100h^{-1}{\rm Mpc}$ (orange solid).
While the ratio with $\pi_{\rm max}=50h^{-1}{\rm Mpc}$ is almost one, the ratio with $\pi_{\rm max}=100h^{-1}{\rm Mpc}$ is $0.94 \pm 0.11$. This is opposite of the bias ratio from lensing ($1.20 \pm 0.12$) and is consistent with the result of \cite{Zu_etal2021} that the low-$M_*$ subsample has a larger bias. However, we must note that the $R$ value and the bias ratios measured with two different integral lengths are all unity within $~1\sigma$ and therefore the projection effects contaminate the bias measurements only mildly. However, as we see hints of subsamples affected by projection with the cross-correlation with spectroscopic data, we do not further explore cross-correlations with photometric galaxies.

%This makes sense.
%Since stellar mass is positively correlated with halo mass, low-$M_*$ clusters reside in low mass halos and they are the ones that are included in the cluster catalog due to the projection effects (i.e., boost in the observed $\lambda$).

%Based on our result, we conclude that stellar mass is strongly correlated with halo mass and therefore cannot be used to detect a halo assembly bias.

\subsection{Subsamples based on other parameters}
\label{sec:res:others}

\begin{figure*}
\centering
\includegraphics[width=0.85\textwidth]{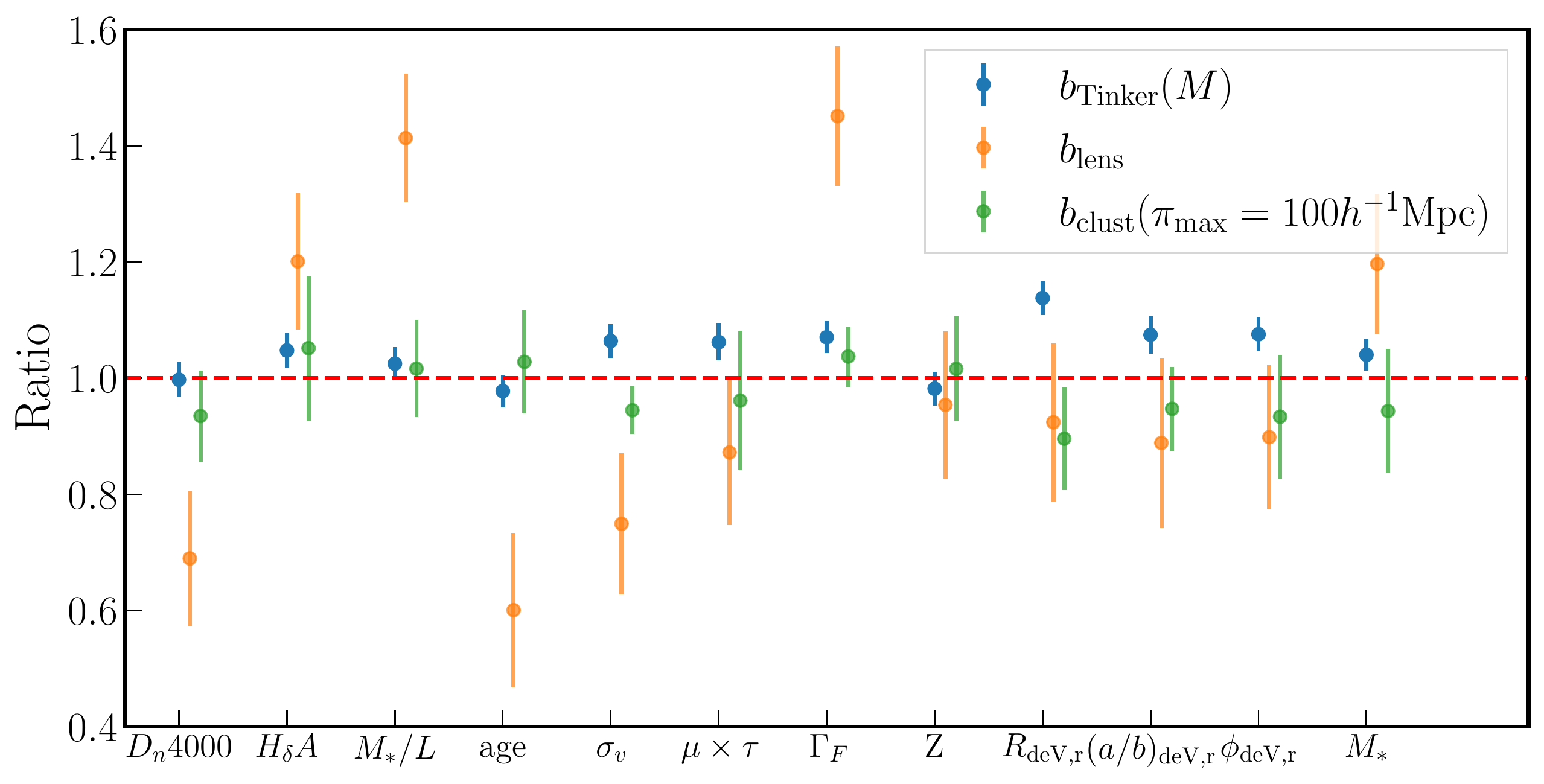}
\caption{\label{fig:bias}Bias ratio for the subsamples split by the central galaxy properties. $b_{\rm Tinker}(M)$ is the predicted halo bias based on halo mass using the Tinker bias model. The symbols $b_{\rm lens}$ and $b_{\rm clust}$ are the halo biases measured from lensing and clustering, respectively. For the bias ratio measured from clustering, we use $\pi_{\rm max}=100\mpch$.}
\end{figure*}

This section presents the results obtained from the cluster subsamples split based on other parameters. The posterior distributions of the model parameters for all the subsamples are summarized in Table~\ref{tab:result_flat}. Fig.~\ref{fig:bias} shows the ratio of the biases of the two subsamples split by different properties. The blue symbols correspond to the theoretical expectation of the ratio of biases given the constraint on halo masses from weak lensing. The orange and green symbols correspond to bias ratios obtained from cluster weak lensing and cross-correlation with spectroscopic galaxies, respectively.

The signal-to-noise ratio of the weak lensing signal is smaller than that of the cross-correlation. Therefore, we primarily compare the green symbols with the blue ones. Overall we see reasonably good agreement between the bias ratios obtained from cross-correlation and that expected from the theoretical expectations, with differences between the two always less than 3$\sigma$. We make some further observations about these results below.

\begin{table*}
	\centering
	\caption{Posterior constraints of the model parameters for the subsamples split on various central galaxy properties listed in Table~\ref{tab:parameter}. The uncertainties are the 1$\sigma$ confidence regions derived from the 1D posterior probability distributions. For these analysis, we use flat priors for all the parameters.}
	\label{tab:result_flat}
	\begin{tabular}{ccccccc} 
		\hline\hline
		Subsample & $M_h[10^{14}\msunh]$ & $c$ & $b$ & $q_{\rm cen}$ & $\alpha_{\rm off}$ & $\chi^2$(dof=15)\\
		\hline
        high-$D_n 4000$ & $1.89^{+0.16}_{-0.14}$  & $4.33^{+0.93}_{-0.68}$  & $2.04^{+0.30}_{-0.30}$  & $0.83^{+0.11}_{-0.12}$  & $0.55^{+0.30}_{-0.34}$   & $11.39$ \\
        low-$D_n 4000$ & $1.90^{+0.11}_{-0.11}$  & $4.69^{+0.70}_{-0.52}$  & $2.96^{+0.30}_{-0.30}$  & $0.76^{+0.18}_{-0.12}$  & $0.10^{+0.27}_{-0.04}$   & $13.97$ \\
        \hline
        high-$H_{\delta}A$ & $2.04^{+0.13}_{-0.12}$  & $4.17^{+0.78}_{-0.49}$  & $2.83^{+0.35}_{-0.36}$  & $0.83^{+0.12}_{-0.16}$  & $0.17^{+0.40}_{-0.11}$   & $12.59$ \\
        low-$H_{\delta}A$ & $1.83^{+0.15}_{-0.13}$  & $4.84^{+0.92}_{-0.70}$  & $2.36^{+0.25}_{-0.24}$  & $0.85^{+0.10}_{-0.11}$  & $0.61^{+0.27}_{-0.36}$   & $27.33$ \\
        \hline
        high-$M_*/L$ & $1.89^{+0.14}_{-0.13}$  & $4.32^{+0.72}_{-0.54}$  & $3.19^{+0.32}_{-0.32}$  & $0.87^{+0.09}_{-0.15}$  & $0.27^{+0.50}_{-0.22}$   & $18.73$ \\
        low-$M_*/L$ & $1.78^{+0.12}_{-0.11}$  & $4.62^{+1.00}_{-0.61}$  & $2.26^{+0.29}_{-0.30}$  & $0.84^{+0.12}_{-0.16}$  & $0.20^{+0.44}_{-0.12}$   & $26.29$ \\
        \hline
        high-age & $1.81^{+0.11}_{-0.10}$  & $5.30^{+0.90}_{-0.64}$  & $1.91^{+0.40}_{-0.39}$  & $0.87^{+0.09}_{-0.14}$  & $0.30^{+0.48}_{-0.23}$   & $14.59$ \\
        low-age & $1.91^{+0.14}_{-0.12}$  & $3.52^{+0.60}_{-0.45}$  & $3.19^{+0.29}_{-0.29}$  & $0.85^{+0.10}_{-0.16}$  & $0.25^{+0.51}_{-0.19}$   & $15.76$ \\
        \hline
        high-$\sigma_v$ & $2.07^{+0.14}_{-0.13}$  & $4.95^{+0.89}_{-0.64}$  & $2.13^{+0.31}_{-0.31}$  & $0.86^{+0.09}_{-0.13}$  & $0.34^{+0.44}_{-0.25}$   & $16.75$ \\
        low-$\sigma_v$ & $1.79^{+0.12}_{-0.11}$  & $4.39^{+0.85}_{-0.60}$  & $2.85^{+0.30}_{-0.30}$  & $0.86^{+0.10}_{-0.13}$  & $0.45^{+0.37}_{-0.35}$   & $14.09$ \\
        \hline
        high-$\mu \times \tau$ & $2.09^{+0.16}_{-0.14}$  & $4.20^{+0.95}_{-0.61}$  & $2.23^{+0.27}_{-0.28}$  & $0.83^{+0.11}_{-0.13}$  & $0.46^{+0.36}_{-0.32}$   & $18.48$ \\
        low-$\mu \times \tau$ & $1.81^{+0.13}_{-0.12}$  & $4.54^{+0.82}_{-0.57}$  & $2.56^{+0.33}_{-0.32}$  & $0.86^{+0.10}_{-0.14}$  & $0.39^{+0.41}_{-0.32}$   & $12.43$ \\
        \hline
        high-$\Gamma_F$ & $2.06^{+0.13}_{-0.12}$  & $3.75^{+0.63}_{-0.44}$  & $3.14^{+0.35}_{-0.35}$  & $0.82^{+0.13}_{-0.15}$  & $0.14^{+0.41}_{-0.08}$   & $16.83$ \\
        low-$\Gamma_F$ & $1.76^{+0.12}_{-0.11}$  & $4.98^{+0.88}_{-0.65}$  & $2.16^{+0.29}_{-0.30}$  & $0.87^{+0.09}_{-0.12}$  & $0.47^{+0.36}_{-0.37}$   & $27.09$ \\
        \hline
        high-Z & $1.88^{+0.13}_{-0.12}$  & $3.35^{+0.63}_{-0.43}$  & $2.50^{+0.27}_{-0.27}$  & $0.85^{+0.11}_{-0.15}$  & $0.31^{+0.46}_{-0.24}$   & $13.28$ \\
        low-Z & $1.96^{+0.14}_{-0.12}$  & $5.29^{+0.91}_{-0.65}$  & $2.62^{+0.37}_{-0.37}$  & $0.86^{+0.10}_{-0.16}$  & $0.23^{+0.50}_{-0.16}$   & $21.10$ \\
        \hline
        high-$R_{\rm deV,r}$ & $2.19^{+0.15}_{-0.13}$  & $4.97^{+0.86}_{-0.54}$  & $2.29^{+0.34}_{-0.34}$  & $0.87^{+0.09}_{-0.14}$  & $0.32^{+0.43}_{-0.24}$   & $15.24$ \\
        low-$R_{\rm deV,r}$ & $1.61^{+0.12}_{-0.11}$  & $3.52^{+0.59}_{-0.46}$  & $2.48^{+0.31}_{-0.31}$  & $0.80^{+0.14}_{-0.15}$  & $0.13^{+0.45}_{-0.07}$   & $8.80$ \\
        \hline
        high-$(a/b)_{\rm deV,r}$ & $2.12^{+0.17}_{-0.15}$  & $3.87^{+0.83}_{-0.57}$  & $2.19^{+0.36}_{-0.36}$  & $0.84^{+0.10}_{-0.12}$  & $0.56^{+0.30}_{-0.36}$   & $8.79$ \\
        low-$(a/b)_{\rm deV,r}$ & $1.79^{+0.13}_{-0.12}$  & $5.28^{+0.94}_{-0.62}$  & $2.47^{+0.32}_{-0.31}$  & $0.86^{+0.10}_{-0.15}$  & $0.27^{+0.47}_{-0.19}$   & $11.16$ \\
        \hline
        high-$\phi_{\rm deV,r}$ & $2.10^{+0.16}_{-0.14}$  & $4.46^{+0.94}_{-0.59}$  & $2.34^{+0.32}_{-0.32}$  & $0.86^{+0.10}_{-0.14}$  & $0.36^{+0.41}_{-0.24}$   & $5.36$ \\
        low-$\phi_{\rm deV,r}$ & $1.77^{+0.10}_{-0.10}$  & $4.83^{+0.87}_{-0.67}$  & $2.61^{+0.30}_{-0.29}$  & $0.73^{+0.19}_{-0.10}$  & $0.11^{+0.13}_{-0.05}$   & $20.22$ \\
        \hline
        high-$M_*$ & $2.01^{+0.13}_{-0.12}$  & $5.31^{+1.01}_{-0.61}$  & $2.74^{+0.32}_{-0.32}$  & $0.85^{+0.10}_{-0.14}$  & $0.28^{+0.46}_{-0.19}$   & $13.63$ \\
        low-$M_*$ & $1.83^{+0.12}_{-0.12}$  & $3.28^{+0.60}_{-0.45}$  & $2.29^{+0.29}_{-0.30}$  & $0.80^{+0.14}_{-0.15}$  & $0.14^{+0.43}_{-0.08}$   & $14.12$ \\
		\hline \hline
	\end{tabular}
	
\end{table*}

Among the 12 parameters we tested for splitting the subsamples, ten resulted in halo masses for the subsamples, consistent within 2$\sigma$. The parameters $D_n 4000$ split gives the closest match in the central values of the posterior distribution of masses for the two subsamples with ${\rm log}_{10}M_h^{\rm high}=14.28^{+0.04}_{-0.03} \msunh$. The ratio of the large scale biases based on lensing is $b_{\rm high}/b_{\rm low}=0.69 \pm 0.12$. We will discuss the case with $D_n 4000$ separately below.

%The rest of the central galaxy properties give different halo masses for the corresponding subsamples, which implies that these parameters are correlated with halo masses. 
%However, the strengths of the correlation with halo masses are different for different parameters.

The parameter $R_{\rm deV,r}$, the size of the central galaxy in the $r$-band, results in a comparatively large difference in halo masses for the two subsamples. The clusters with larger BCGs at fixed richness result in a subsample with ${\rm log}_{10}M_h^{\rm high}=14.34\pm 0.03 \msunh$, while that with smaller BCGs have halo mass ${\rm log}_{10}M_h^{\rm low}=14.21 \pm 0.03 \msunh$. Albeit at smaller significance, this difference is consistent with the recent study by \cite{song_etal2021} that the outer profiles of massive galaxies are sensitive to halo mass even at fixed richness.

\begin{table*}
	\centering
	\caption{Bias ratio for the subsamples split by the central galaxy properties. $b_{\rm Tinker}$ is the predicted halo bias based on halo mass using the Tinker bias model. The columns $b_{\rm lens}$ and $b_{\rm clust}$ are the halo biases measured from lensing and clustering, respectively. For the bias ratio measured from clustering, we use two different integral scales $\pi_{\rm max}=50\mpch$ and $100\mpch$.}
	\label{tab:bias}
	\begin{tabular}{ccccc} 
		\hline\hline
		Parameter & $b_{\rm Tinker}$ & $b_{\rm lens}$ & $b_{\rm clust}(\pi_{\rm max}=50\mpch)$ & $b_{\rm clust}(\pi_{\rm max}=100\mpch)$ \\
		\hline
		$D_n 4000$ & $1.00^{+0.03}_{-0.03}$  & $0.69^{+0.12}_{-0.12}$  & $0.94^{+0.03}_{-0.03}$   & $0.93^{+0.08}_{-0.08}$ \\
        $H_{\delta}A$ & $1.05^{+0.03}_{-0.03}$  & $1.20^{+0.12}_{-0.12}$  & $1.02^{+0.10}_{-0.10}$   & $1.05^{+0.12}_{-0.12}$ \\
        $M_*/L$ & $1.02^{+0.03}_{-0.03}$  & $1.41^{+0.11}_{-0.11}$  & $0.95^{+0.03}_{-0.03}$   & $1.02^{+0.08}_{-0.08}$ \\
        age & $0.98^{+0.03}_{-0.03}$  & $0.60^{+0.13}_{-0.13}$  & $0.97^{+0.04}_{-0.04}$   & $1.03^{+0.09}_{-0.09}$ \\
        $\sigma_v$ & $1.06^{+0.03}_{-0.03}$  & $0.75^{+0.12}_{-0.12}$  & $0.98^{+0.05}_{-0.05}$   & $0.94^{+0.04}_{-0.04}$ \\
        $\mu \times \tau$ & $1.06^{+0.03}_{-0.03}$  & $0.87^{+0.13}_{-0.13}$  & $1.02^{+0.11}_{-0.11}$   & $0.96^{+0.12}_{-0.12}$ \\
        $\Gamma_F$ & $1.07^{+0.03}_{-0.03}$  & $1.45^{+0.12}_{-0.12}$  & $1.01^{+0.03}_{-0.03}$   & $1.04^{+0.05}_{-0.05}$ \\
        Z & $0.98^{+0.03}_{-0.03}$  & $0.95^{+0.13}_{-0.13}$  & $1.00^{+0.04}_{-0.04}$   & $1.02^{+0.09}_{-0.09}$ \\
        $R_{\rm deV,r}$ & $1.14^{+0.03}_{-0.03}$  & $0.92^{+0.14}_{-0.14}$  & $0.96^{+0.05}_{-0.05}$   & $0.90^{+0.09}_{-0.09}$ \\
        $(a/b)_{\rm deV,r}$ & $1.07^{+0.03}_{-0.03}$  & $0.89^{+0.15}_{-0.14}$  & $0.98^{+0.05}_{-0.05}$   & $0.95^{+0.07}_{-0.07}$ \\
        $\phi_{\rm deV,r}$ & $1.08^{+0.03}_{-0.03}$  & $0.90^{+0.12}_{-0.12}$  & $0.96^{+0.08}_{-0.08}$   & $0.93^{+0.11}_{-0.11}$ \\
        $M_*$ & $1.04^{+0.03}_{-0.03}$  & $1.20^{+0.12}_{-0.12}$  & $1.01^{+0.07}_{-0.07}$   & $0.94^{+0.11}_{-0.11}$ \\
        \hline \hline

	\end{tabular}
	
\end{table*}

% Given the lensing errors, I would think this is not worth mentioning much!
%%%% Among all the parameters except $D_n 4000$, the subsamples split by $H \delta_A$, $\Gamma_F$, $age$, $M_*$, and $M_*/L$ show consistent relations between halo masses and halo biases (i.e., the subsample with a larger halo mass has a larger halo bias than the other subsample). 
%%%% However, other parameters ($(a/b)_{\rm deV,r}$, $\phi_{\rm deV,r}$, $R_{\rm deV,r}$, $\mu \times \tau$, and $\sigma_v$) exhibit an opposite trend that the subsample with a larger halo mass has a smaller halo bias.
%%%% Possible causes of this are halo assembly bias and the projection effects.
%%%% To make sure that the cause is not the projection effects, we measure the clustering ratio $w_{\rm p,cg}^{\rm high}(R)/w_{\rm p,cg}^{\rm low}(R)$ at $\pi_{\rm max}=50h^{-1}{\rm Mpc}$ and $100h^{-1}{\rm Mpc}$. The bias ratios measured from clustering are listed in Table~\ref{tab:bias} and in Fig.~\ref{fig:bias}.
%--------------------------------------------

We do not observe significant projection effects among all the parameters we consider. Since central galaxy properties are unlikely to be contaminated by interlopers, we expect the contamination due to the projection effects should not be significant.%This implies that any of these five parameters are possibly correlated with halo assembly history besides halo masses.

For parameters which show significant differences in halo masses, one could iterate over the sample split criteria and use the weak lensing to ensure subsamples have equal halo mass \citep[see, e.g.,][]{Lin2016}. We leave such investigations for future work.

\begin{table*}
	\centering
	\caption{Bias ratio for the subsamples split by the central galaxy properties. $b_{\rm lens}$ and $b_{\rm clust}$ are the halo biases measured from lensing and clustering, respectively. For the bias ratio measured from clustering, we use several different scales to fit. Note that the unit for $R$ is $[\mpch]$.}
	\label{tab:bias_scale}
	\begin{tabular}{ccccc} 
		\hline\hline
		Parameter & $b_{\rm lens}$ & $b_{\rm clust}:R \in [10,50]$ & $b_{\rm clust}:R \in [5,50]$ & $b_{\rm clust}:R \in [2,50]$ \\
		\hline
		$M_*$ & $1.20^{+0.12}_{-0.12}$  & $0.94^{+0.11}_{-0.11}$   & $0.98^{+0.09}_{-0.09}$   & $0.91^{+0.09}_{-0.09}$ \\
        \hline
        $D_n 4000$ & $0.69^{+0.12}_{-0.12}$  & $0.93^{+0.08}_{-0.08}$   & $0.91^{+0.08}_{-0.08}$   & $0.89^{+0.08}_{-0.08}$ \\
        \hline
        $M_*/L$ & $1.41^{+0.11}_{-0.11}$  & $1.02^{+0.08}_{-0.08}$   & $1.10^{+0.08}_{-0.08}$   & $1.04^{+0.08}_{-0.08}$ \\
        \hline
        age & $0.60^{+0.13}_{-0.13}$  & $1.03^{+0.09}_{-0.09}$   & $0.94^{+0.09}_{-0.09}$   & $0.96^{+0.09}_{-0.09}$ \\
        \hline
        $\Gamma_F$ & $1.45^{+0.12}_{-0.12}$  & $1.04^{+0.05}_{-0.05}$   & $1.12^{+0.06}_{-0.06}$   & $1.06^{+0.07}_{-0.07}$ \\
       
		\hline \hline
	\end{tabular}
	
\end{table*}

%Furthermore, we compare the bias ratios measured from lensing and clustering. 
%Most of parameters give a consistent results for these bias ratios. 

Finally, we also comment on the consistency of bias ratios of sub-samples between lensing and clustering, which appear consistent for most of the properties. The parameters $M_*/L$, $age$ and $\Gamma_F$ show that the bias ratio measured from clustering is almost one, while the bias ratio measured from lensing for these parameters shows more than $30\%$ gap between the two subsamples. It is not clear what causes this difference. We cross-check if this difference could result from the choice of scale for clustering. We measure the bias ratio on several different scales using the clustering as shown in Table~\ref{tab:bias_scale}. However, we do not see significant differences in the biases measured from the different scales. %Our original measurement is done on $10\mpch < R < 50 \mpch$, and we additionally measure the bias ratio on $5\mpch < R < 50 \mpch$ and  $2\mpch < R < 50 \mpch$. 
%Interestingly, the bias ratio measured on  $5\mpch < R < 50 \mpch$ has the closest value to the bias ratio from lensing.
%However, the change was less than $10\%$, and the choice of the scale cannot fully explain the discrepancy between the bias ratios from lensing and clustering.

Another possibility is our choice of cosmology; the amplitude of lensing can be lower than that of clustering by assuming a \textit{Planck} cosmology \citep{amon_etal2022}. The difference between lensing and clustering amplitudes can be $\sim 20\%$. However, if this inconsistency persists on both subsamples, taking the ratio should cancel out. We could potentially investigate this issue using shape measurements from the Hyper Suprime Cam SSP (HSC) survey \citep{li_etal2022} due to its deeper photometry and better image quality than the SDSS source catalogue. We select all the source galaxies behind the lens galaxies in the current work. However, due to the shallow depth of SDSS, we cannot completely rule out source-cluster member confusion. We will leave the investigation of such effects to future work.

%We primarily want to measure the halo mass and bias via the weak lensing measurements. If the measured halo masses for these subsamples are the same but biases are different, it can be a possible detection of halo assembly bias and we want to proceed to make sure that the bias difference between the two subsamples is not due to the projection effects via cluster-galaxy cross-correlation.
%However, it is often difficult to split the cluster sample into equal masses in the first place, since we cannot know the halo mass of each cluster before the split.
%One way to mitigate this issue is to iterate the sample split and the weak lensing mass measurement until the two subsamples have equal halo mass, which is used in \cite{Lin2016}.
%Instead, we decide to use the expected halo bias inferred from the measured halo mass of the cluster sample using the halo bias modeled by \cite{tinker10} from \textit{Colossus} \citep{Diemer2018}, when the two subsamples do not have equal masses.
%By comparing the measured bias ratio and the expected bias ratio predicted from the halo masses, we can tell whether the measured bias difference can be explained by the halo mass difference of the two subsamples.
%If the measured bias difference is different from the expected value, it means that a bias difference between the subsamples cannot be fully explained by halo mass and the parameter used to split the cluster sample is possibly correlated with halo assembly history, though the fact that the two subsamples do not have the equal mass means that the parameter is also correlated with halo mass.

\begin{figure*}
\centering
\includegraphics[width=0.8\textwidth]{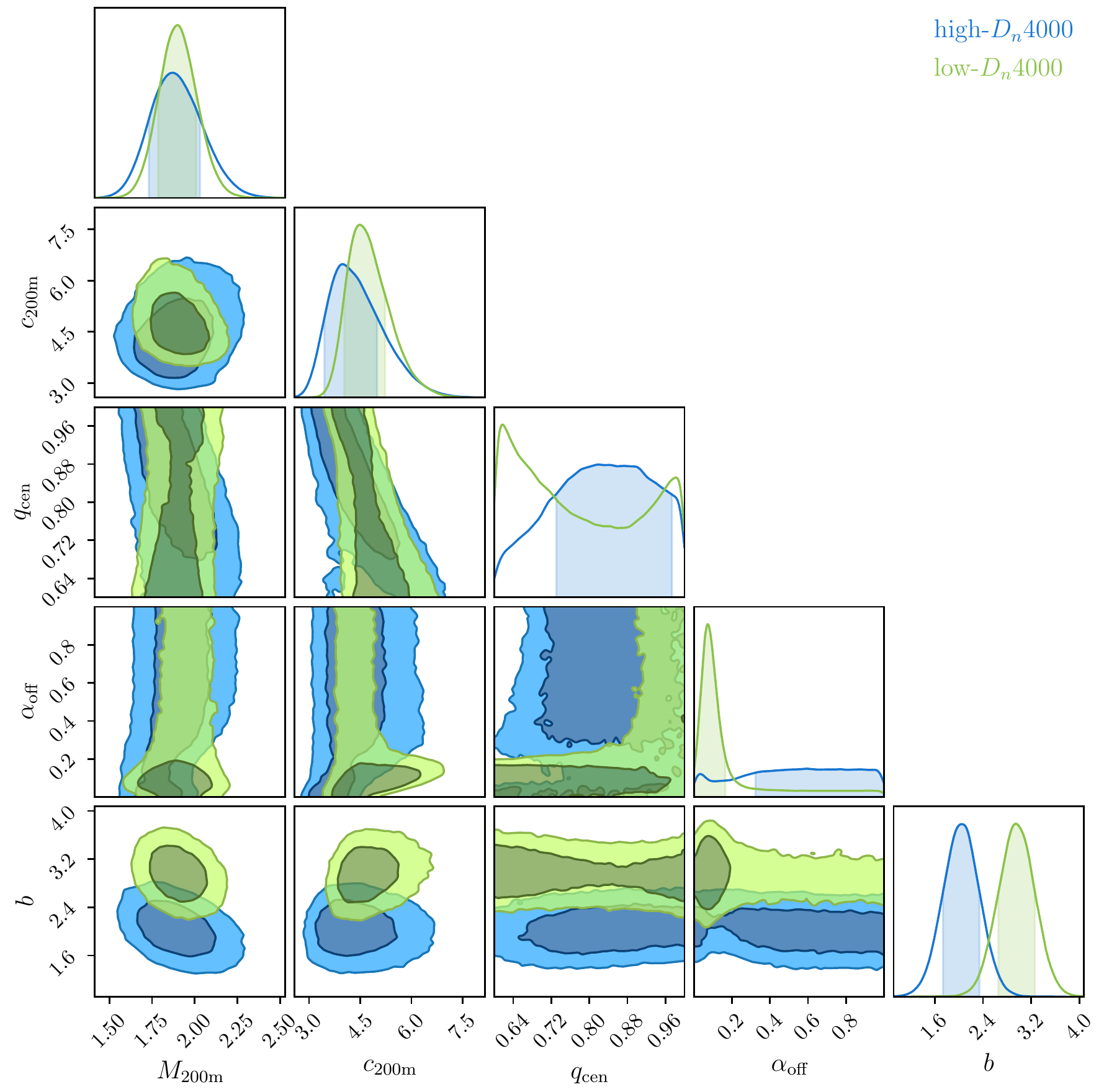}
\caption{\label{fig:triangle_dn4000}  Posterior constraint from the modelling of $\Delta \Sigma(R)$ for the high-$D_n 4000$ (blue) and low-$D_n 4000$ (green) subsamples. Each histogram in the diagonal panels are the 1D marginalised posterior distribution of the parameters, while each contour plot in the off-diagonal panels indicate the 1$\sigma$ and 2$\sigma$ confidence regions of the matching parameter pairs.}
\end{figure*}

Finally, we focus on the results for the subsamples split by $D_n 4000$. Fig.~\ref{fig:triangle_dn4000} shows the derived posterior constraints of our model parameters for the high-$D_n 4000$ (blue) and low-$D_n 4000$ (green) subsamples, similar to Fig.~\ref{fig:triangle_mstar}.
The parameter $D_n 4000$, which is a measure of the flux break at the 4000 $\angstrom$ break strength, is sensitive to the average of the stellar population and is a good tracer of the long-term star formation histories (e.g., \cite{paulino-afonso_etal2020}).
The mean halo masses of the high-$D_n 4000$ and low-$D_n 4000$ subsamples are almost equal, and the corresponding halo biases from lensing for these subsamples are different by $\sim 40\%$ (although the bias ratio is consistent with unity at less than 3 $\sigma$).

At last, we compare the bias ratio measured from lensing and clustering. While the bias ratio from lensing is $\sim 40\%$, the one from clustering is $\sim 5\%$. Therefore, we cannot plausibly claim that subsamples split by $D_n 4000$ show halo assembly bias.
%, and we need further investigation to accurately measure the halo bias ratio as well as the concentration parameter, which are crucial components in the detection of halo assembly bias.
%{\bf Snip text until here.}

%\begin{figure}
%\centering
%\includegraphics[width=0.45\textwidth]{wp_size.png}
%\caption{\label{fig:ror_size} Top: Cluster-galaxy projected cross-correlation functions integrated up to $\pi_{\rm max}=100h^{-1}{\rm Mpc}$ for the large/small-$R_{\rm deV,r}$ subsamples. Middle:The ratio $w_{\rm p,cg}^{\rm high}(R)$ and $w_{\rm p,cg}^{\rm low}(R)$ computed with $\pi_{\rm max}=50h^{-1}{\rm Mpc}$ (blue dash-dot) and $\pi_{\rm max}=100h^{-1}{\rm Mpc}$ (orange solid). Bottom: The ratio of $w_{\rm p,cg}^{\rm high}(R)/w_{\rm p,cg}^{\rm low}(R)$ at $\pi_{\rm max}=50h^{-1}{\rm Mpc}$ and $\pi_{\rm max}=100h^{-1}{\rm Mpc}$ denoted as "Ratio of Ratio".}
%\end{figure}

%\section{Discussion: clustering and lensing}
%In this section, we discuss some examples that the measured biases from lensing signals and clustering differ largely. 
%4) age is a bit extreme in halo bias, m/L
%5)clustering to lensing　measured biases are different: Fracstar, HdeltaA, age, M/L
%We should compare the clustering and lensing signals and how the measured biases can differ. Also, discuss the choice of cosmology for the measurements.
%1. bias from clustering vs. bias from lensing
%2. the effect of the choice of cosmology

\section{Conclusions}

Halo assembly bias is the correlation between the large scale clustering of halos and secondary properties of the halos at fixed mass. It has been studied extensively using numerical simulations but has been difficult to confirm in observations. In this paper, we have searched for signs of halo assembly bias based on the central galaxy properties of SDSS redMaPPer clusters. Our research work can be summarized as follows:

%We split the cluster sample into two by using the properties of their central galaxies at fixed richness and redshift. We tested twelve different central galaxy properties listed in Table~\ref{tab:parameter}. To avoid the spectroscopic incompleteness of the redMaPPer clusters pointed out by \cite{Zu_etal2021}, we only use the cluster central galaxies with $M_i \leq -22.4$, where the detection fraction by spectroscopic observation is complete.

%Using these clusters, we measured cluster weak lensing profiles $\Delta \Sigma(R)$ and cluster-galaxy cross-correlation functions $w_{\rm p,cg}(R)$ for the two subsamples. To avoid the false detection of halo assembly bias due to the projection effects, we measured the bias ratio through $w_{\rm p,cg}(R)$ with $\pi_{\rm max}=50h^{-1}{\rm Mpc}$ and $100 h^{-1}{\rm Mpc}$. This analysis step is a guardrail against bias differences due to the projection effects.

%We summarize our conclusions as follows:
\begin{itemize}
\item We used 12 different properties of the central galaxies in SDSS redMaPPer clusters derived from the analysis of their SDSS spectra by \citet{Chen_etal2012}. We subdivided redMaPPer clusters using the median of the relationship between these parameters and the cluster richness.
\item We measured the weak lensing signals for each cluster sub-samples and modelled it to infer the average halo mass of these sub-samples and their large scale biases from the cluster weak lensing signal.
\item We also measured the cross-correlations between galaxy clusters and spectroscopic galaxies from SDSS to obtain a higher signal-to-noise ratio estimate for the large scale clustering.
\item For sub-samples obtained by using ten out of the twelve properties we considered, the halo masses are consistent with each other within 2$\sigma$. We computed the theoretically expected ratios of biases based on the halo mass estimates for all sub-sample splits.
\item The halo bias ratios of each sub-samples as obtained by the cross-correlations and the expectations based on halo mass difference are consistent with each other.
%\item Among the twelve parameters we tested, $D_n 4000$ is the only parameter, which did not show the correlation with halo masses (i.e., the inferred halo masses for the high/low-$D_n 4000$ subsamples are almost equal, ${\rm log}_{10}M_h^{high/low}=14.29 \msunh$).
%\item The bias ratio for the high/low-$D_n 4000$ subsamples from lensing is $b_{\rm high}/b_{\rm low}=0.69 \pm 0.12$, while that from clustering is $b_{\rm high}/b_{\rm low}=0.93 \pm 0.08$.
%\item We found that stellar mass and halo mass are positively correlated at fixed $\lambda$, consistent with the SHMR prediction.
%\item Among the twelve parameters we tested, the size of the central galaxy correlates the most with halo mass.
\item We do not see significant evidence of contamination by projection effects in sub-sample splits based on central galaxy properties. %The two subsamples which are unequally affected by the projection effects at 1$\sigma$ level are stellar mass $M_*$ and size $R_{\rm deV,r}$.
\end{itemize}

In addition to the above findings, we also find some cases, albeit at low significance, where the bias ratio measured from the lensing signal differs from the ratio measured from clustering for some parameters.
%Even for the case of $D_n 4000$, the bias ratios from lensing and clustering were different, and we do not conclude that this is a detection of halo assembly bias.
Such inconsistencies between lensing and clustering signals require further investigation as they could potentially signal issues in the contamination of clusters. In future work, we will use the shape catalogue from the HSC survey \citep{li_etal2022}, which overlaps with the footprint of the redMaPPer clusters. %The HSC data has deeper photometry and better image quality than the SDSS source catalog and enables us to further investigate the effects of the source galaxy selections such as intrinsic alignments. 
%We plan to explore the inconsistency between lensing and clustering signals as our future follow-up studies.

\section*{Acknowledgements}
We thank Eduardo Rozo and Elisabeth Krause for their suggestions. 
TS is supported by Grant-in-Aid for JSPS Fellows 20J01600 and JSPS KAKENHI Grant Number 20H05855.
HM is supported by JSPS KAKENHI Grant Numbers JP21H05456.
%Investigator Award AS-IA-109-M02 

Funding for SDSS-III has been provided by the Alfred P. Sloan Foundation, the Participating Institutions, the National Science Foundation, and the U.S. Department of Energy Office of Science. The SDSS-III web site is \url{http://www.sdss3.org/.}

SDSS-III is managed by the Astrophysical Research Consortium for the Participating Institutions of the SDSS-III Collaboration including the University of Arizona, the Brazilian Participation Group, Brookhaven National Laboratory, Carnegie Mellon University, University of Florida, the French Participation Group, the German Participation Group, Harvard University, the Instituto de Astrofisica de Canarias, the Michigan State/Notre Dame/JINA Participation Group, Johns Hopkins University, Lawrence Berkeley National Laboratory, Max Planck Institute for Astrophysics, Max Planck Institute for Extraterrestrial Physics, New Mexico State University, New York University, Ohio State University, Pennsylvania State University, University of Portsmouth, Princeton University, the Spanish Participation Group, University of Tokyo, University of Utah, Vanderbilt University, University of Virginia, University of Washington, and Yale University.

\section*{Data Availability}
The SDSS redMaPPer cluster v5.10 catalogs are publicly available at http://risa.stanford.edu/redmapper/, and BOSS DR12 LOWZ galaxy sample is available at https://data.sdss.org/sas/dr12/boss/lss/.
The lensing and clustering measurements are currently not publicly available, but the measurements are available upon request from the authors.

%%%%%%%%%%%%%%%%% APPENDICES %%%%%%%%%%%%%%%%%%%%%

\appendix

\section{Appendix}
\subsection{Clusters without luminosity cut}
\label{app:no_lum}

\begin{table*}
	\centering
	\caption{Posterior constraints of the model parameters for the subsamples split by stellar masses $M_*$. We use the redMaPPer clusters without luminosity cut (i.e., include more low-luminosity clusters in the redMaPPer cluster catalog).}
	\label{tab:result_nolum}
	\begin{tabular}{ccccccc} 
		\hline\hline
		Subsample & $M_h[10^{14}\msunh]$ & $c$ & $b$ & $q_{\rm cen}$ & $\alpha_{\rm off}$ & $\chi^2$ (dof=15) \\
		\hline \hline
        high-$M_*$ & $1.98^{+0.13}_{-0.12}$  & $5.41^{+1.12}_{-0.66}$  & $2.64^{+0.29}_{-0.30}$  & $0.84^{+0.11}_{-0.15}$  & $0.25^{+0.45}_{-0.16}$   & $10.41$ \\
        low-$M_*$ & $1.73^{+0.11}_{-0.11}$  & $3.39^{+0.55}_{-0.44}$  & $2.54^{+0.31}_{-0.32}$  & $0.78^{+0.16}_{-0.13}$  & $0.13^{+0.36}_{-0.07}$   & $10.80$ \\
        \hline \hline
	\end{tabular}
	
\end{table*}

For the results presented above, we use the redMaPPer clusters with spectroscopic follow-up at $M_i \leq -22.4$ to avoid a possible spectroscopic incompleteness. 
We examine whether this luminosity cut affects the results significantly. For this purpose, we repeat the same analysis but without using a luminosity cut.

We show the posterior constraints of our model parameters for the high-$M_*$ and low-$M_*$ subsamples inferred from our analysis in Table.~\ref{tab:result_nolum}. 
As expected, the mean halo masses for these subsamples are lower by $\sim10\%$ than those for the subsamples with luminosity cut.
This result is because the subsamples without luminosity cut include more low luminous central galaxies likely to reside in lower-mass halos.
Similarly, the halo biases are smaller by $\sim10\%$, and the rest of the parameters do not change as much as the halo mass and bias.

\subsection{Results for the cluster subsamples with $p_{\rm cen}\geq 0.9$}
\label{app:pcen}
In this subsection, we repeat the same analysis for the cluster subsamples with the central probability $p_{\rm cen}\geq0.9$. The inferred posterior constraints of our model parameters for the subsamples are shown in Table.~\ref{tab:result_pcen}. For these analysis, we changed the prior range for $q_{\rm cen}$ to $q_{\rm cen} \sim \mathcal{U}(0.9,1.)$.

\begin{table*}
	\centering
	\caption{Posterior constraints of the model parameters for the subsamples split by stellar masses $M_*$ and $D_n 4000$ with $p_{\rm cen}\geq0.9$. }
	\label{tab:result_pcen}
	\begin{tabular}{ccccccc} 
		\hline\hline
		Subsample & $M_h[10^{14}\msunh]$ & $c$ & $b$ & $q_{\rm cen}$ & $\alpha_{\rm off}$ & $\chi^2$ (dof=15) \\
		\hline \hline
        high-$M_*$ & $2.04^{+0.13}_{-0.13}$  & $4.91^{+0.50}_{-0.45}$  & $2.59^{+0.36}_{-0.36}$  & $0.95^{+0.04}_{-0.03}$  & $0.55^{+0.31}_{-0.35}$   & $20.12$ \\
        low-$M_*$ & $1.79^{+0.14}_{-0.14}$  & $3.57^{+0.45}_{-0.41}$  & $2.26^{+0.34}_{-0.34}$  & $0.95^{+0.03}_{-0.03}$  & $0.51^{+0.34}_{-0.34}$   & $9.33$ \\
        \hline
        high-$D_n 4000$ & $1.82^{+0.15}_{-0.15}$  & $4.26^{+0.58}_{-0.51}$  & $1.98^{+0.37}_{-0.37}$  & $0.94^{+0.04}_{-0.03}$  & $0.55^{+0.31}_{-0.34}$   & $17.65$ \\
        low-$D_n 4000$ & $1.85^{+0.16}_{-0.16}$  & $4.50^{+0.52}_{-0.49}$  & $2.87^{+0.39}_{-0.38}$  & $0.95^{+0.03}_{-0.03}$  & $0.48^{+0.35}_{-0.33}$   & $22.58$ \\
        \hline \hline
	\end{tabular}
	
\end{table*}

%In this subsection, we repeat the same analysis as described in Sec.~\ref{sec:lensing}, but with flat priors for all five parameters. For our fiducial analysis, and thus for the results presented above, we used the Gaussian priors for the concentration and two off-centering parameters to break the degeneracy among these parameters. 
%We examine how the results shift if we use flat priors for these parameters, and use the flat priors for $x \sim \mathcal{U}(1, 10)$, $q \sim \mathcal{U}(0.6, 1.0)$, and $\alpha_{\rm off} \sim \mathcal{U}(0.0001,1.)$.

%The derived posterior constraints of our model parameters for all the subsamples are shown in Table.~\ref{tab:result_flat}. 
%As is expected, the halo mass and bias are not affected by this change. However, the value of the fraction of centered clusters $q_{\rm cen}$ shifts from $\sim 0.7$ to $\sim 0.8$. This is because we used the Gaussian priors of $\mathcal{N}(0.3,0.042)$ in our original analysis. Among all the clusters used in our analysis, the fraction of the clusters whose centering probability $p_{\rm cen} \geq 0.8$ is roughly $\sim 76\%$. So, getting a larger $q_{\rm cen}$ by using the flat prior can be interpreted to be consistent with the fraction of clusters with $p_{\rm cen} \geq 0.8$.
%The value of concentration, on the other hand, becomes smaller by $\sim 10\%$ on average, and the variation of $\alpha_{\rm off}$ becomes larger and goes to $\sim 0.6$ for some subsamples.

\bibliographystyle{Frontiers-Harvard}
\bibliography{central}

%%%%%%%%%

% Don't change these lines
%\section*{Table captions}

%\section*{Figure captions}

% if your bibtex file is called example.bib{}

%\bsp	% typesetting comment
% Don't change these lines
\bsp	% typesetting comment
\label{lastpage}
\end{document}

\onecolumn
\firstpage{1}

\title[Supplementary Material]{{\helveticaitalic{Supplementary Material}}}

\maketitle

%\section{Supplementary Data}

%Supplementary Material should be uploaded separately on submission. Please include any supplementary data, figures and/or tables. All supplementary files are deposited to FigShare for permanent storage and receive a DOI.

%Supplementary material is not typeset so please ensure that all information is clearly presented, the appropriate caption is included in the file and not in the manuscript, and that the style conforms to the rest of the article. To avoid discrepancies between the published article and the supplementary material, please do not add the title, author list, affiliations or correspondence in the supplementary files.

\section{Supplementary Tables}

%\subsection{Tables}

\begin{table}[h!]
	\centering
	\caption{List of central galaxy properties used to find the correlation with halo assembly history.}
	\label{tab:parameter}
    \scalebox{0.85}{
	\begin{tabular}{cc} 
		\hline\hline
		Parameter & Description \\
		\hline
		${\rm D}_{n}4000$ & the 4000$\angstrom$ break strength\\
		$H{\delta}_{A}$ & the strength of Balmer absorption line\\
		$M_{*}/L$ & estimates of stellar mass-to-light ratio\\
		age & the r-band luminosity-weighted age\\
		$\mu \times \tau$ & the dust parameters $\mu$ and $\tau$\\
		$\sigma_v$ & the stellar velocity dispersion\\
		 $\Gamma_F$ &  fraction of stars formed in recent bursts \\
		 $Z$ & metallicity \\
		 $R_{{\rm deV},r}$ & de Vaucouleurs fit scale radius in $r$ band\\
		 $(a/b)_{{\rm deV},r}$ &  de Vaucouleurs fit a/b in $r$ band\\
		 $\phi_{{\rm deV},r}$ & de Vaucouleurs fit position angle in $r$ band\\
		$M_{*}$ & stellar mass\\
		\hline \hline
	\end{tabular}}
\end{table}

\begin{table}[h!]
	\centering
	\caption{Posterior constraints of the model parameters for the subsamples split on various central galaxy properties listed in Table~\ref{tab:parameter}. The uncertainties are the 1$\sigma$ confidence regions derived from the 1D posterior probability distributions.}
	\label{tab:result}
	\scalebox{0.85}{
	\begin{tabular}{ccccccc} 
		\hline\hline
		Subsample & $M_h[10^{14}\msunh]$ & $c$ & $b$ & $q_{\rm cen}$ & $\alpha_{\rm off}$ & $\chi^2$(dof=15)) \\
        \hline
		high-$D_n 4000$ & $1.93^{+0.16}_{-0.15}$  & $4.83^{+0.76}_{-0.66}$  & $2.24^{+0.36}_{-0.35}$  & $0.72^{+0.09}_{-0.07}$  & $0.29^{+0.05}_{-0.06}$   & $19.81$ \\
        low-$D_n 4000$ & $1.94^{+0.12}_{-0.12}$  & $4.95^{+0.72}_{-0.61}$  & $3.09^{+0.37}_{-0.37}$  & $0.69^{+0.09}_{-0.07}$  & $0.07^{+0.05}_{-0.04}$   & $19.50$ \\
        \hline
        high-$H_{\delta}A$ & $2.10^{+0.15}_{-0.14}$  & $4.74^{+0.78}_{-0.66}$  & $2.95^{+0.41}_{-0.42}$  & $0.69^{+0.09}_{-0.06}$  & $0.14^{+0.12}_{-0.09}$   & $13.04$ \\
        low-$H_{\delta}A$ & $1.82^{+0.15}_{-0.14}$  & $5.26^{+0.96}_{-0.86}$  & $2.35^{+0.30}_{-0.32}$  & $0.75^{+0.09}_{-0.08}$  & $0.21^{+0.10}_{-0.13}$   & $27.80$ \\
        \hline
        high-$M_*/L$ & $1.94^{+0.17}_{-0.15}$  & $4.82^{+0.77}_{-0.67}$  & $3.21^{+0.38}_{-0.41}$  & $0.72^{+0.10}_{-0.08}$  & $0.18^{+0.10}_{-0.09}$   & $18.34$ \\
        low-$M_*/L$ & $1.77^{+0.12}_{-0.12}$  & $5.56^{+0.86}_{-0.84}$  & $2.44^{+0.34}_{-0.36}$  & $0.70^{+0.09}_{-0.07}$  & $0.16^{+0.09}_{-0.08}$   & $32.72$ \\
        \hline
        high-age & $1.81^{+0.12}_{-0.11}$  & $5.60^{+0.92}_{-0.84}$  & $1.99^{+0.43}_{-0.44}$  & $0.73^{+0.10}_{-0.08}$  & $0.14^{+0.14}_{-0.10}$   & $18.81$ \\
        low-age & $2.00^{+0.16}_{-0.15}$  & $4.26^{+0.69}_{-0.65}$  & $3.38^{+0.35}_{-0.38}$  & $0.70^{+0.09}_{-0.07}$  & $0.21^{+0.11}_{-0.14}$   & $20.71$ \\
        \hline
        high-$\sigma_v$ & $2.00^{+0.15}_{-0.14}$  & $4.93^{+1.00}_{-0.79}$  & $2.60^{+0.36}_{-0.37}$  & $0.70^{+0.09}_{-0.07}$  & $0.16^{+0.14}_{-0.11}$   & $30.41$ \\
        low-$\sigma_v$ & $1.88^{+0.17}_{-0.16}$  & $4.65^{+0.90}_{-0.72}$  & $2.88^{+0.36}_{-0.38}$  & $0.73^{+0.09}_{-0.08}$  & $0.20^{+0.13}_{-0.15}$   & $18.30$ \\
        \hline
        high-$\mu \times \tau$ & $2.12^{+0.16}_{-0.15}$  & $4.91^{+0.81}_{-0.73}$  & $2.37^{+0.34}_{-0.33}$  & $0.71^{+0.09}_{-0.07}$  & $0.26^{+0.07}_{-0.12}$   & $20.99$ \\
        low-$\mu \times \tau$ & $1.83^{+0.14}_{-0.13}$  & $4.81^{+0.78}_{-0.64}$  & $2.64^{+0.38}_{-0.39}$  & $0.71^{+0.10}_{-0.07}$  & $0.09^{+0.12}_{-0.05}$   & $12.58$ \\
        \hline
        high-$\Gamma_F$ & $2.09^{+0.15}_{-0.14}$  & $4.33^{+0.66}_{-0.56}$  & $3.27^{+0.43}_{-0.47}$  & $0.68^{+0.10}_{-0.05}$  & $0.12^{+0.10}_{-0.06}$   & $22.05$ \\
        low-$\Gamma_F$ & $1.73^{+0.13}_{-0.12}$  & $5.31^{+0.88}_{-0.83}$  & $2.25^{+0.36}_{-0.35}$  & $0.76^{+0.09}_{-0.09}$  & $0.23^{+0.08}_{-0.11}$   & $32.07$ \\
        \hline
        high-Z & $1.92^{+0.15}_{-0.14}$  & $4.23^{+0.71}_{-0.66}$  & $2.55^{+0.31}_{-0.34}$  & $0.70^{+0.09}_{-0.07}$  & $0.22^{+0.09}_{-0.12}$   & $18.82$ \\
        low-Z & $1.99^{+0.14}_{-0.13}$  & $5.49^{+0.87}_{-0.80}$  & $2.78^{+0.44}_{-0.45}$  & $0.72^{+0.10}_{-0.08}$  & $0.13^{+0.12}_{-0.07}$   & $23.59$ \\
        \hline
        high-$R_{\rm deV,r}$ & $2.17^{+0.15}_{-0.14}$  & $5.84^{+0.82}_{-0.79}$  & $2.31^{+0.38}_{-0.38}$  & $0.77^{+0.09}_{-0.09}$  & $0.22^{+0.07}_{-0.10}$   & $20.61$ \\
        low-$R_{\rm deV,r}$ & $1.64^{+0.15}_{-0.13}$  & $4.03^{+0.64}_{-0.60}$  & $2.56^{+0.38}_{-0.39}$  & $0.66^{+0.09}_{-0.05}$  & $0.13^{+0.09}_{-0.05}$   & $10.49$ \\
        \hline
        high-$(a/b)_{\rm deV,r}$ & $2.22^{+0.17}_{-0.17}$  & $4.54^{+0.78}_{-0.66}$  & $2.10^{+0.41}_{-0.40}$  & $0.72^{+0.09}_{-0.08}$  & $0.29^{+0.06}_{-0.12}$   & $14.95$ \\
        low-$(a/b)_{\rm deV,r}$ & $1.77^{+0.13}_{-0.12}$  & $5.44^{+0.90}_{-0.71}$  & $2.55^{+0.38}_{-0.39}$  & $0.71^{+0.10}_{-0.07}$  & $0.07^{+0.10}_{-0.05}$   & $13.81$ \\
        \hline
        high-$\phi_{\rm deV,r}$ & $2.12^{+0.16}_{-0.15}$  & $5.29^{+0.82}_{-0.75}$  & $2.40^{+0.38}_{-0.39}$  & $0.74^{+0.09}_{-0.08}$  & $0.25^{+0.06}_{-0.06}$   & $11.25$ \\
        low-$\phi_{\rm deV,r}$ & $1.77^{+0.12}_{-0.11}$  & $4.94^{+0.76}_{-0.68}$  & $2.73^{+0.34}_{-0.35}$  & $0.68^{+0.09}_{-0.05}$  & $0.10^{+0.05}_{-0.04}$   & $17.03$ \\
        \hline
        high-$M_*$ & $2.06^{+0.14}_{-0.13}$  & $5.96^{+0.87}_{-0.86}$  & $2.74^{+0.36}_{-0.41}$  & $0.76^{+0.09}_{-0.08}$  & $0.22^{+0.08}_{-0.10}$   & $17.84$ \\
        low-$M_*$ & $1.82^{+0.15}_{-0.13}$  & $4.02^{+0.65}_{-0.61}$  & $2.50^{+0.36}_{-0.37}$  & $0.67^{+0.09}_{-0.05}$  & $0.13^{+0.09}_{-0.06}$   & $15.32$ \\
		\hline \hline
	\end{tabular}
	}
\end{table}

\begin{table}[h!]
	\centering
	\caption{Bias ratio for the subsamples split by the central galaxy properties. $b_{\rm Tinker}$ is the predicted halo bias based on halo mass using the Tinker bias model, while $b_{\rm lens}$ and $b_{\rm clust}$ are the halo biases measured from lensing and clustering, respectively. For the bias ratio measured from clustering, we use two different integral scales $\pi_{\rm max}=50\mpch$ and $100\mpch$.}
	\label{tab:bias}
	\scalebox{0.85}{
	\begin{tabular}{ccccc} 
		\hline\hline
		Parameter & $b_{\rm Tinker}$ & $b_{\rm lens}$ & $b_{\rm clust}(\pi_{\rm max}=50\mpch)$ & $b_{\rm clust}(\pi_{\rm max}=100\mpch)$ \\
		\hline
		$D_n 4000$ & $1.00^{+0.03}_{-0.03}$  & $0.69^{+0.12}_{-0.12}$  & $0.94^{+0.03}_{-0.03}$   & $0.93^{+0.08}_{-0.08}$ \\
        $H_{\delta}A$ & $1.05^{+0.03}_{-0.03}$  & $1.20^{+0.12}_{-0.12}$  & $1.02^{+0.10}_{-0.10}$   & $1.05^{+0.12}_{-0.12}$ \\
        $M_*/L$ & $1.02^{+0.03}_{-0.03}$  & $1.41^{+0.11}_{-0.11}$  & $0.95^{+0.03}_{-0.03}$   & $1.02^{+0.08}_{-0.08}$ \\
        age & $0.98^{+0.03}_{-0.03}$  & $0.60^{+0.13}_{-0.13}$  & $0.97^{+0.04}_{-0.04}$   & $1.03^{+0.09}_{-0.09}$ \\
        $\sigma_v$ & $1.06^{+0.03}_{-0.03}$  & $0.75^{+0.12}_{-0.12}$  & $0.98^{+0.05}_{-0.05}$   & $0.94^{+0.04}_{-0.04}$ \\
        $\mu \times \tau$ & $1.06^{+0.03}_{-0.03}$  & $0.87^{+0.13}_{-0.13}$  & $1.02^{+0.11}_{-0.11}$   & $0.96^{+0.12}_{-0.12}$ \\
        $\Gamma_F$ & $1.07^{+0.03}_{-0.03}$  & $1.45^{+0.12}_{-0.12}$  & $1.01^{+0.03}_{-0.03}$   & $1.04^{+0.05}_{-0.05}$ \\
        Z & $0.98^{+0.03}_{-0.03}$  & $0.95^{+0.13}_{-0.13}$  & $1.00^{+0.04}_{-0.04}$   & $1.02^{+0.09}_{-0.09}$ \\
        $R_{\rm deV,r}$ & $1.14^{+0.03}_{-0.03}$  & $0.92^{+0.14}_{-0.14}$  & $0.96^{+0.05}_{-0.05}$   & $0.90^{+0.09}_{-0.09}$ \\
        $(a/b)_{\rm deV,r}$ & $1.07^{+0.03}_{-0.03}$  & $0.89^{+0.15}_{-0.14}$  & $0.98^{+0.05}_{-0.05}$   & $0.95^{+0.07}_{-0.07}$ \\
        $\phi_{\rm deV,r}$ & $1.08^{+0.03}_{-0.03}$  & $0.90^{+0.12}_{-0.12}$  & $0.96^{+0.08}_{-0.08}$   & $0.93^{+0.11}_{-0.11}$ \\
        $M_*$ & $1.04^{+0.03}_{-0.03}$  & $1.20^{+0.12}_{-0.12}$  & $1.01^{+0.07}_{-0.07}$   & $0.94^{+0.11}_{-0.11}$ \\
        \hline \hline

	\end{tabular}
	}
\end{table}
\begin{table}
	\centering
	\caption{List of central galaxy properties used to find the correlation with halo assembly history.}
	\label{tab:parameter}
    \scalebox{0.85}{
	\begin{tabular}{cc} 
		\hline\hline
		Parameter & Description \\
		\hline
		${\rm D}_{n}4000$ & the 4000$\angstrom$ break strength\\
		$H{\delta}_{A}$ & the strength of Balmer absorption line\\
		$M_{*}/L$ & estimates of stellar mass-to-light ratio\\
		age & the r-band luminosity-weighted age\\
		$\mu \times \tau$ & the dust parameters $\mu$ and $\tau$\\
		$\sigma_v$ & the stellar velocity dispersion\\
		 $\Gamma_F$ &  fraction of stars formed in recent bursts \\
		 $Z$ & metallicity \\
		 $R_{{\rm deV},r}$ & de Vaucouleurs fit scale radius in $r$ band\\
		 $(a/b)_{{\rm deV},r}$ &  de Vaucouleurs fit a/b in $r$ band\\
		 $\phi_{{\rm deV},r}$ & de Vaucouleurs fit position angle in $r$ band\\
		$M_{*}$ & stellar mass\\
		\hline \hline
	\end{tabular}}
\end{table}

\begin{table}
	\centering
	\caption{Posterior constraints of the model parameters for the subsamples split on various central galaxy properties listed in Table~\ref{tab:parameter}. The uncertainties are the 1$\sigma$ confidence regions derived from the 1D posterior probability distributions. For these analysis, we use flat priors for all the parameters.}
	\label{tab:result_flat}
	\scalebox{0.85}{
	\begin{tabular}{ccccccc} 
		\hline\hline
		Subsample & $M_h[10^{14}\msunh]$ & $c$ & $b$ & $q_{\rm cen}$ & $\alpha_{\rm off}$ & $\chi^2$(dof=15)\\
		\hline
        high-$D_n 4000$ & $1.89^{+0.16}_{-0.14}$  & $4.33^{+0.93}_{-0.68}$  & $2.04^{+0.30}_{-0.30}$  & $0.83^{+0.11}_{-0.12}$  & $0.55^{+0.30}_{-0.34}$   & $11.39$ \\
        low-$D_n 4000$ & $1.90^{+0.11}_{-0.11}$  & $4.69^{+0.70}_{-0.52}$  & $2.96^{+0.30}_{-0.30}$  & $0.76^{+0.18}_{-0.12}$  & $0.10^{+0.27}_{-0.04}$   & $13.97$ \\
        \hline
        high-$H_{\delta}A$ & $2.04^{+0.13}_{-0.12}$  & $4.17^{+0.78}_{-0.49}$  & $2.83^{+0.35}_{-0.36}$  & $0.83^{+0.12}_{-0.16}$  & $0.17^{+0.40}_{-0.11}$   & $12.59$ \\
        low-$H_{\delta}A$ & $1.83^{+0.15}_{-0.13}$  & $4.84^{+0.92}_{-0.70}$  & $2.36^{+0.25}_{-0.24}$  & $0.85^{+0.10}_{-0.11}$  & $0.61^{+0.27}_{-0.36}$   & $27.33$ \\
        \hline
        high-$M_*/L$ & $1.89^{+0.14}_{-0.13}$  & $4.32^{+0.72}_{-0.54}$  & $3.19^{+0.32}_{-0.32}$  & $0.87^{+0.09}_{-0.15}$  & $0.27^{+0.50}_{-0.22}$   & $18.73$ \\
        low-$M_*/L$ & $1.78^{+0.12}_{-0.11}$  & $4.62^{+1.00}_{-0.61}$  & $2.26^{+0.29}_{-0.30}$  & $0.84^{+0.12}_{-0.16}$  & $0.20^{+0.44}_{-0.12}$   & $26.29$ \\
        \hline
        high-age & $1.81^{+0.11}_{-0.10}$  & $5.30^{+0.90}_{-0.64}$  & $1.91^{+0.40}_{-0.39}$  & $0.87^{+0.09}_{-0.14}$  & $0.30^{+0.48}_{-0.23}$   & $14.59$ \\
        low-age & $1.91^{+0.14}_{-0.12}$  & $3.52^{+0.60}_{-0.45}$  & $3.19^{+0.29}_{-0.29}$  & $0.85^{+0.10}_{-0.16}$  & $0.25^{+0.51}_{-0.19}$   & $15.76$ \\
        \hline
        high-$\sigma_v$ & $2.07^{+0.14}_{-0.13}$  & $4.95^{+0.89}_{-0.64}$  & $2.13^{+0.31}_{-0.31}$  & $0.86^{+0.09}_{-0.13}$  & $0.34^{+0.44}_{-0.25}$   & $16.75$ \\
        low-$\sigma_v$ & $1.79^{+0.12}_{-0.11}$  & $4.39^{+0.85}_{-0.60}$  & $2.85^{+0.30}_{-0.30}$  & $0.86^{+0.10}_{-0.13}$  & $0.45^{+0.37}_{-0.35}$   & $14.09$ \\
        \hline
        high-$\mu \times \tau$ & $2.09^{+0.16}_{-0.14}$  & $4.20^{+0.95}_{-0.61}$  & $2.23^{+0.27}_{-0.28}$  & $0.83^{+0.11}_{-0.13}$  & $0.46^{+0.36}_{-0.32}$   & $18.48$ \\
        low-$\mu \times \tau$ & $1.81^{+0.13}_{-0.12}$  & $4.54^{+0.82}_{-0.57}$  & $2.56^{+0.33}_{-0.32}$  & $0.86^{+0.10}_{-0.14}$  & $0.39^{+0.41}_{-0.32}$   & $12.43$ \\
        \hline
        high-$\Gamma_F$ & $2.06^{+0.13}_{-0.12}$  & $3.75^{+0.63}_{-0.44}$  & $3.14^{+0.35}_{-0.35}$  & $0.82^{+0.13}_{-0.15}$  & $0.14^{+0.41}_{-0.08}$   & $16.83$ \\
        low-$\Gamma_F$ & $1.76^{+0.12}_{-0.11}$  & $4.98^{+0.88}_{-0.65}$  & $2.16^{+0.29}_{-0.30}$  & $0.87^{+0.09}_{-0.12}$  & $0.47^{+0.36}_{-0.37}$   & $27.09$ \\
        \hline
        high-Z & $1.88^{+0.13}_{-0.12}$  & $3.35^{+0.63}_{-0.43}$  & $2.50^{+0.27}_{-0.27}$  & $0.85^{+0.11}_{-0.15}$  & $0.31^{+0.46}_{-0.24}$   & $13.28$ \\
        low-Z & $1.96^{+0.14}_{-0.12}$  & $5.29^{+0.91}_{-0.65}$  & $2.62^{+0.37}_{-0.37}$  & $0.86^{+0.10}_{-0.16}$  & $0.23^{+0.50}_{-0.16}$   & $21.10$ \\
        \hline
        high-$R_{\rm deV,r}$ & $2.19^{+0.15}_{-0.13}$  & $4.97^{+0.86}_{-0.54}$  & $2.29^{+0.34}_{-0.34}$  & $0.87^{+0.09}_{-0.14}$  & $0.32^{+0.43}_{-0.24}$   & $15.24$ \\
        low-$R_{\rm deV,r}$ & $1.61^{+0.12}_{-0.11}$  & $3.52^{+0.59}_{-0.46}$  & $2.48^{+0.31}_{-0.31}$  & $0.80^{+0.14}_{-0.15}$  & $0.13^{+0.45}_{-0.07}$   & $8.80$ \\
        \hline
        high-$(a/b)_{\rm deV,r}$ & $2.12^{+0.17}_{-0.15}$  & $3.87^{+0.83}_{-0.57}$  & $2.19^{+0.36}_{-0.36}$  & $0.84^{+0.10}_{-0.12}$  & $0.56^{+0.30}_{-0.36}$   & $8.79$ \\
        low-$(a/b)_{\rm deV,r}$ & $1.79^{+0.13}_{-0.12}$  & $5.28^{+0.94}_{-0.62}$  & $2.47^{+0.32}_{-0.31}$  & $0.86^{+0.10}_{-0.15}$  & $0.27^{+0.47}_{-0.19}$   & $11.16$ \\
        \hline
        high-$\phi_{\rm deV,r}$ & $2.10^{+0.16}_{-0.14}$  & $4.46^{+0.94}_{-0.59}$  & $2.34^{+0.32}_{-0.32}$  & $0.86^{+0.10}_{-0.14}$  & $0.36^{+0.41}_{-0.24}$   & $5.36$ \\
        low-$\phi_{\rm deV,r}$ & $1.77^{+0.10}_{-0.10}$  & $4.83^{+0.87}_{-0.67}$  & $2.61^{+0.30}_{-0.29}$  & $0.73^{+0.19}_{-0.10}$  & $0.11^{+0.13}_{-0.05}$   & $20.22$ \\
        \hline
        high-$M_*$ & $2.01^{+0.13}_{-0.12}$  & $5.31^{+1.01}_{-0.61}$  & $2.74^{+0.32}_{-0.32}$  & $0.85^{+0.10}_{-0.14}$  & $0.28^{+0.46}_{-0.19}$   & $13.63$ \\
        low-$M_*$ & $1.83^{+0.12}_{-0.12}$  & $3.28^{+0.60}_{-0.45}$  & $2.29^{+0.29}_{-0.30}$  & $0.80^{+0.14}_{-0.15}$  & $0.14^{+0.43}_{-0.08}$   & $14.12$ \\
		\hline \hline
	\end{tabular}
	}
\end{table}

\begin{table}
	\centering
	\caption{Bias ratio for the subsamples split by the central galaxy properties. $b_{\rm Tinker}$ is the predicted halo bias based on halo mass using the Tinker bias model. The columns $b_{\rm lens}$ and $b_{\rm clust}$ are the halo biases measured from lensing and clustering, respectively. For the bias ratio measured from clustering, we use two different integral scales $\pi_{\rm max}=50\mpch$ and $100\mpch$.}
	\label{tab:bias}
	\scalebox{0.85}{
	\begin{tabular}{ccccc} 
		\hline\hline
		Parameter & $b_{\rm Tinker}$ & $b_{\rm lens}$ & $b_{\rm clust}(\pi_{\rm max}=50\mpch)$ & $b_{\rm clust}(\pi_{\rm max}=100\mpch)$ \\
		\hline
		$D_n 4000$ & $1.00^{+0.03}_{-0.03}$  & $0.69^{+0.12}_{-0.12}$  & $0.94^{+0.03}_{-0.03}$   & $0.93^{+0.08}_{-0.08}$ \\
        $H_{\delta}A$ & $1.05^{+0.03}_{-0.03}$  & $1.20^{+0.12}_{-0.12}$  & $1.02^{+0.10}_{-0.10}$   & $1.05^{+0.12}_{-0.12}$ \\
        $M_*/L$ & $1.02^{+0.03}_{-0.03}$  & $1.41^{+0.11}_{-0.11}$  & $0.95^{+0.03}_{-0.03}$   & $1.02^{+0.08}_{-0.08}$ \\
        age & $0.98^{+0.03}_{-0.03}$  & $0.60^{+0.13}_{-0.13}$  & $0.97^{+0.04}_{-0.04}$   & $1.03^{+0.09}_{-0.09}$ \\
        $\sigma_v$ & $1.06^{+0.03}_{-0.03}$  & $0.75^{+0.12}_{-0.12}$  & $0.98^{+0.05}_{-0.05}$   & $0.94^{+0.04}_{-0.04}$ \\
        $\mu \times \tau$ & $1.06^{+0.03}_{-0.03}$  & $0.87^{+0.13}_{-0.13}$  & $1.02^{+0.11}_{-0.11}$   & $0.96^{+0.12}_{-0.12}$ \\
        $\Gamma_F$ & $1.07^{+0.03}_{-0.03}$  & $1.45^{+0.12}_{-0.12}$  & $1.01^{+0.03}_{-0.03}$   & $1.04^{+0.05}_{-0.05}$ \\
        Z & $0.98^{+0.03}_{-0.03}$  & $0.95^{+0.13}_{-0.13}$  & $1.00^{+0.04}_{-0.04}$   & $1.02^{+0.09}_{-0.09}$ \\
        $R_{\rm deV,r}$ & $1.14^{+0.03}_{-0.03}$  & $0.92^{+0.14}_{-0.14}$  & $0.96^{+0.05}_{-0.05}$   & $0.90^{+0.09}_{-0.09}$ \\
        $(a/b)_{\rm deV,r}$ & $1.07^{+0.03}_{-0.03}$  & $0.89^{+0.15}_{-0.14}$  & $0.98^{+0.05}_{-0.05}$   & $0.95^{+0.07}_{-0.07}$ \\
        $\phi_{\rm deV,r}$ & $1.08^{+0.03}_{-0.03}$  & $0.90^{+0.12}_{-0.12}$  & $0.96^{+0.08}_{-0.08}$   & $0.93^{+0.11}_{-0.11}$ \\
        $M_*$ & $1.04^{+0.03}_{-0.03}$  & $1.20^{+0.12}_{-0.12}$  & $1.01^{+0.07}_{-0.07}$   & $0.94^{+0.11}_{-0.11}$ \\
        \hline \hline

	\end{tabular}
	}
\end{table}

\begin{table}[h!]
	\centering
	\caption{Bias ratio for the subsamples split by the central galaxy properties. $b_{\rm lens}$ and $b_{\rm clust}$ are the halo biases measured from lensing and clustering, respectively. For the bias ratio measured from clustering, we use several different scales to fit. Note that the unit for $R$ is $[\mpch]$.}
	\label{tab:bias_scale}
	\scalebox{0.85}{
	\begin{tabular}{ccccc} 
		\hline\hline
		Parameter & $b_{\rm lens}$ & $b_{\rm clust}:R \in [10,50]$ & $b_{\rm clust}:R \in [5,50]$ & $b_{\rm clust}:R \in [2,50]$ \\
		\hline
		$M_*$ & $1.20^{+0.12}_{-0.12}$  & $0.94^{+0.11}_{-0.11}$   & $0.98^{+0.09}_{-0.09}$   & $0.91^{+0.09}_{-0.09}$ \\
        \hline
        $D_n 4000$ & $0.69^{+0.12}_{-0.12}$  & $0.93^{+0.08}_{-0.08}$   & $0.91^{+0.08}_{-0.08}$   & $0.89^{+0.08}_{-0.08}$ \\
        \hline
        $M_*/L$ & $1.41^{+0.11}_{-0.11}$  & $1.02^{+0.08}_{-0.08}$   & $1.10^{+0.08}_{-0.08}$   & $1.04^{+0.08}_{-0.08}$ \\
        \hline
        age & $0.60^{+0.13}_{-0.13}$  & $1.03^{+0.09}_{-0.09}$   & $0.94^{+0.09}_{-0.09}$   & $0.96^{+0.09}_{-0.09}$ \\
        \hline
        $\Gamma_F$ & $1.45^{+0.12}_{-0.12}$  & $1.04^{+0.05}_{-0.05}$   & $1.12^{+0.06}_{-0.06}$   & $1.06^{+0.07}_{-0.07}$ \\
        \hline
		\hline \hline
	\end{tabular}
	}
\end{table}

\begin{table}[h!]
	\centering
	\caption{Posterior constraints of the model parameters for the subsamples split by stellar masses $M_*$. We use the redMaPPer clusters without luminosity cut (i.e., include more low-luminosity clusters in the redMaPPer cluster catalog).}
	\label{tab:result_nolum}
	\scalebox{0.85}{
	\begin{tabular}{ccccccc} 
		\hline\hline
		Subsample & $M_h[10^{14}\msunh]$ & $c$ & $b$ & $q_{\rm cen}$ & $\alpha_{\rm off}$ & $\chi^2$ (dof=15) \\
		\hline \hline
        high-$M_*$ & $1.98^{+0.13}_{-0.12}$  & $5.41^{+1.12}_{-0.66}$  & $2.64^{+0.29}_{-0.30}$  & $0.84^{+0.11}_{-0.15}$  & $0.25^{+0.45}_{-0.16}$   & $10.41$ \\
        low-$M_*$ & $1.73^{+0.11}_{-0.11}$  & $3.39^{+0.55}_{-0.44}$  & $2.54^{+0.31}_{-0.32}$  & $0.78^{+0.16}_{-0.13}$  & $0.13^{+0.36}_{-0.07}$   & $10.80$ \\
        \hline \hline
	\end{tabular}
	}
\end{table}

\begin{table}[h!]
	\centering
	\caption{Posterior constraints of the model parameters for the subsamples split by stellar masses $M_*$ and $D_n 4000$ with $p_{\rm cen}\geq0.9$. }
	\label{tab:result_pcen}
	\scalebox{0.85}{
	\begin{tabular}{ccccccc} 
		\hline\hline
		Subsample & $M_h[10^{14}\msunh]$ & $c$ & $b$ & $q_{\rm cen}$ & $\alpha_{\rm off}$ & $\chi^2$ (dof=15) \\
		\hline \hline
        high-$M_*$ & $2.04^{+0.13}_{-0.13}$  & $4.91^{+0.50}_{-0.45}$  & $2.59^{+0.36}_{-0.36}$  & $0.95^{+0.04}_{-0.03}$  & $0.55^{+0.31}_{-0.35}$   & $20.12$ \\
        low-$M_*$ & $1.79^{+0.14}_{-0.14}$  & $3.57^{+0.45}_{-0.41}$  & $2.26^{+0.34}_{-0.34}$  & $0.95^{+0.03}_{-0.03}$  & $0.51^{+0.34}_{-0.34}$   & $9.33$ \\
        \hline
        high-$D_n 4000$ & $1.82^{+0.15}_{-0.15}$  & $4.26^{+0.58}_{-0.51}$  & $1.98^{+0.37}_{-0.37}$  & $0.94^{+0.04}_{-0.03}$  & $0.55^{+0.31}_{-0.34}$   & $17.65$ \\
        low-$D_n 4000$ & $1.85^{+0.16}_{-0.16}$  & $4.50^{+0.52}_{-0.49}$  & $2.87^{+0.39}_{-0.38}$  & $0.95^{+0.03}_{-0.03}$  & $0.48^{+0.35}_{-0.33}$   & $22.58$ \\
        \hline \hline
	\end{tabular}
	}
\end{table}

\section{Supplementary Figures}

\begin{figure}[h!]
\centering
\includegraphics[width=0.45\textwidth]{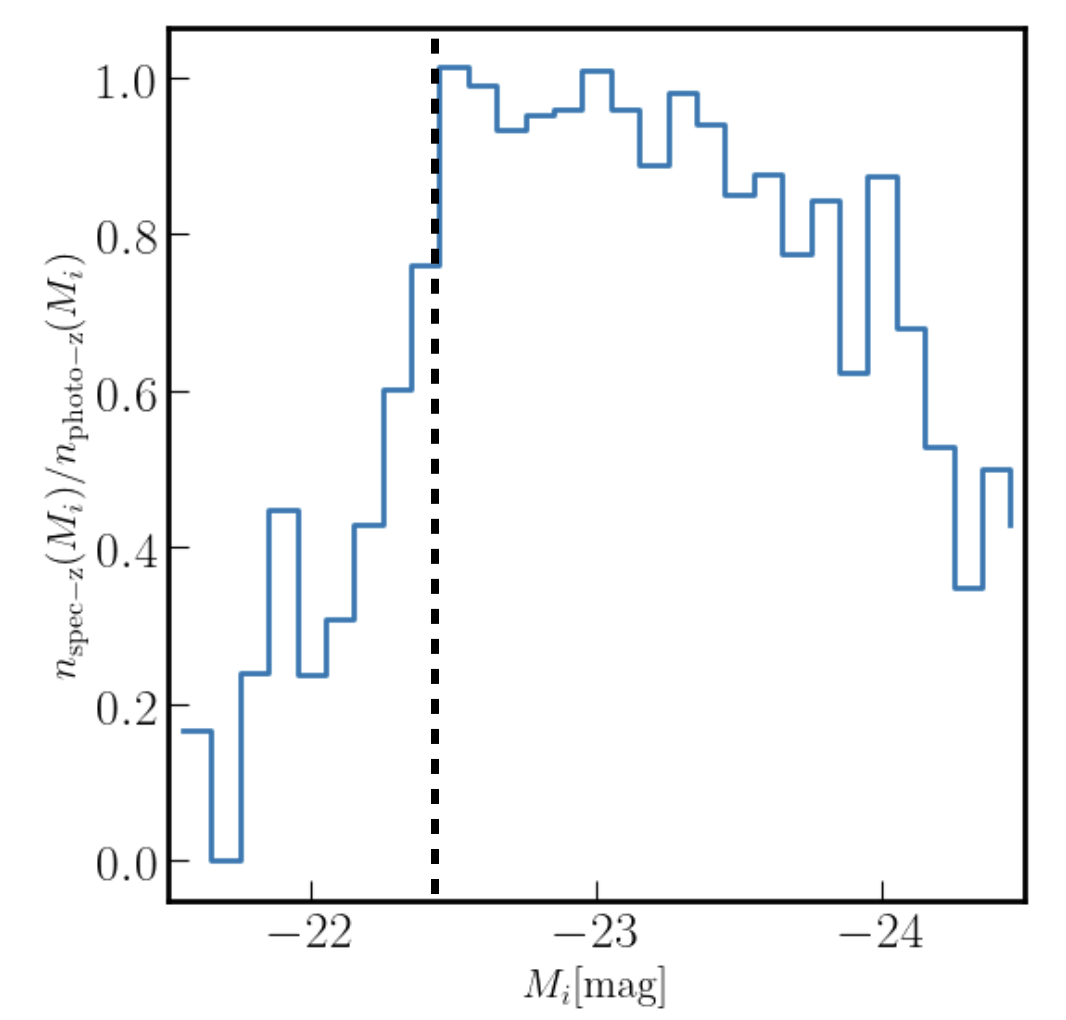}
\caption{\label{fig:mcut} We select redMaPPer clusters whose absolute magnitude is $M_{i}<-22.4$.}
\end{figure}

\begin{figure}[h!]
\centering
\includegraphics[width=0.95\textwidth]{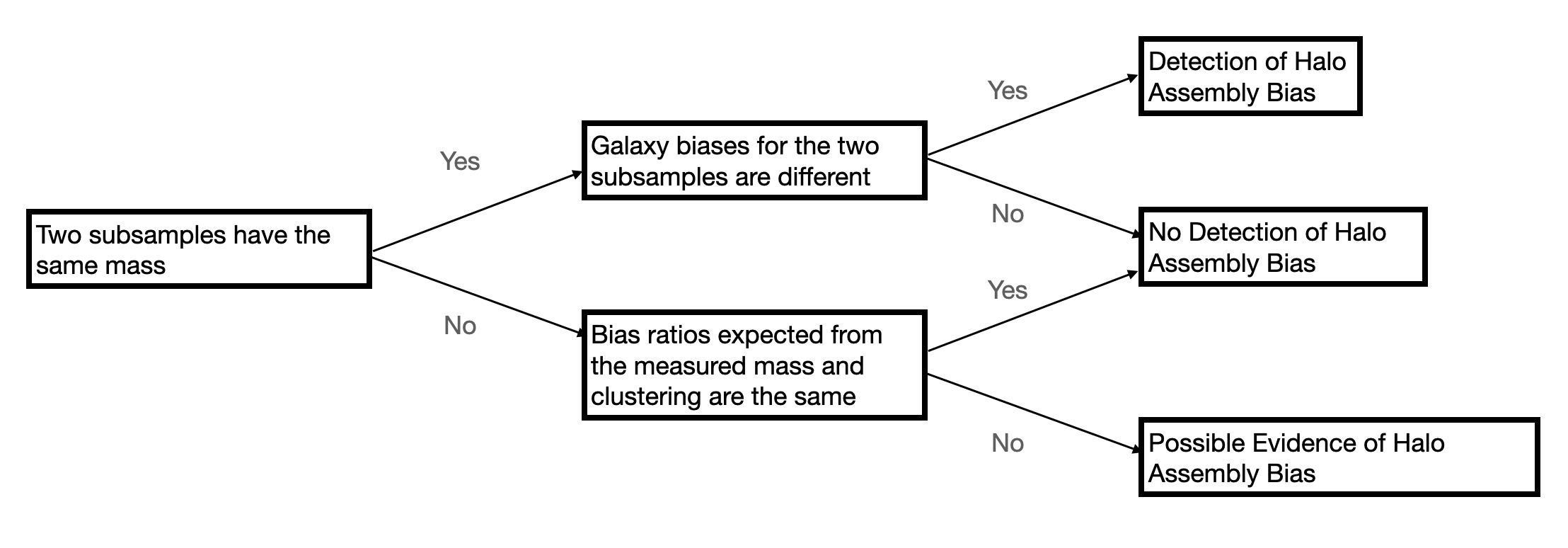}
\caption{\label{fig:logic1} The logic to detect halo assembly bias often used in other studies.}
\end{figure}

\begin{figure}[h!]
\centering
\includegraphics[width=0.95\textwidth]{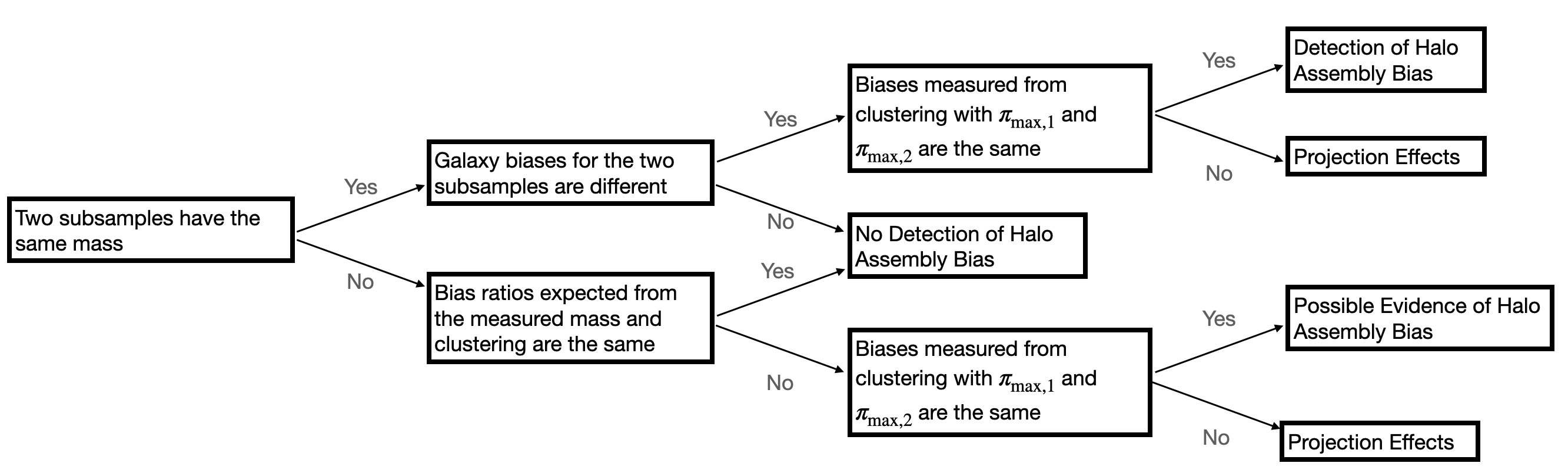}
\caption{\label{fig:logic2} The logic to detect halo assembly bias by decoupling from the projection effects.}
\end{figure}

\begin{figure}[h!]
\centering
\includegraphics[width=0.8\textwidth]{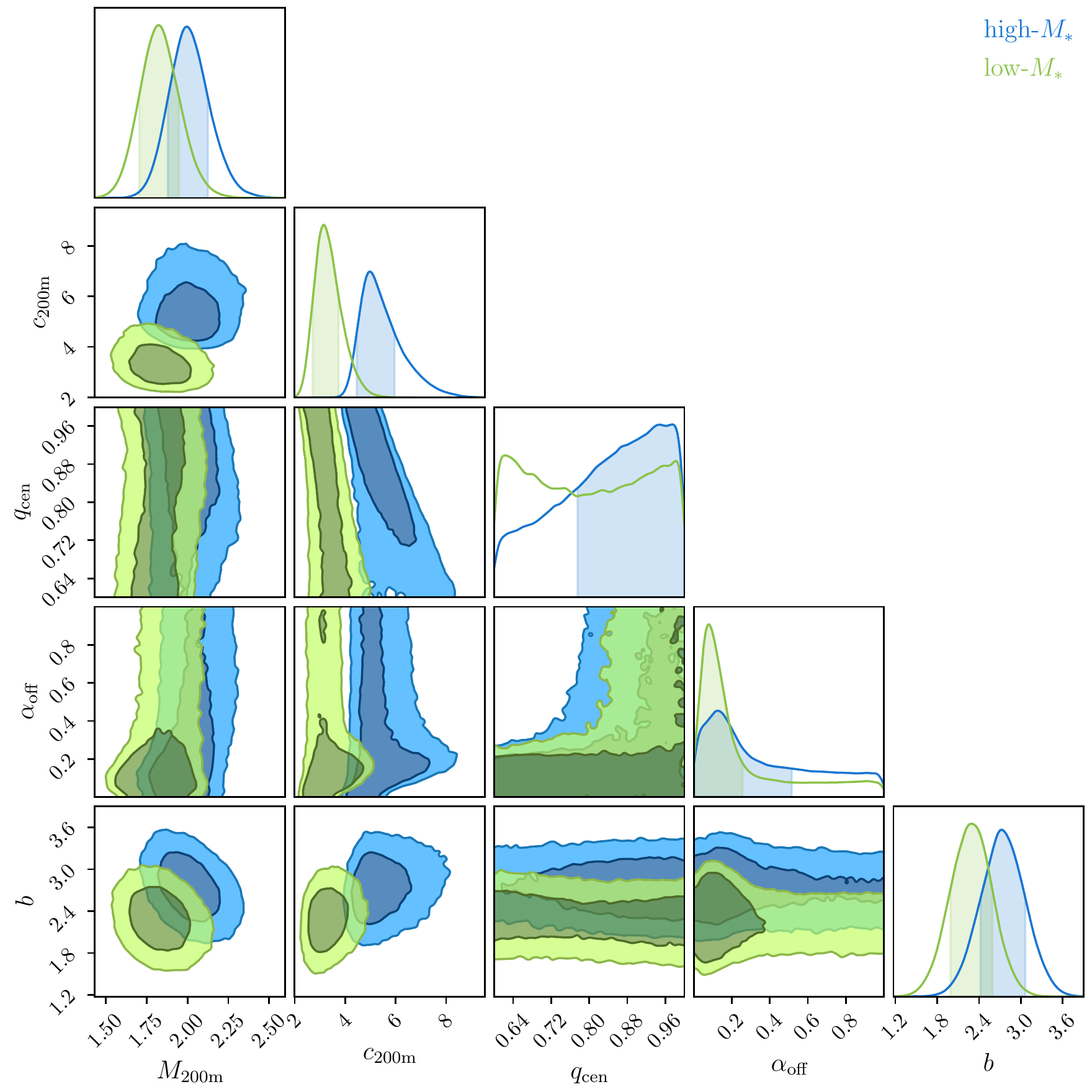}
\caption{\label{fig:triangle_mstar}  Posterior constraint from the modelling of $\Delta \Sigma(R)$ for the high-$M_*$ (blue) and low-$M_*$ (green) subsamples. Each histogram in the diagonal panels are the 1D marginalised posterior distribution of the parameters, while each contour plot in the off-diagonal panels indicate the 1$\sigma$ and 2$\sigma$ confidence regions of the matching parameter pairs.}
\end{figure}

\begin{figure}[h!]
\centering
\includegraphics[width=0.45\textwidth]{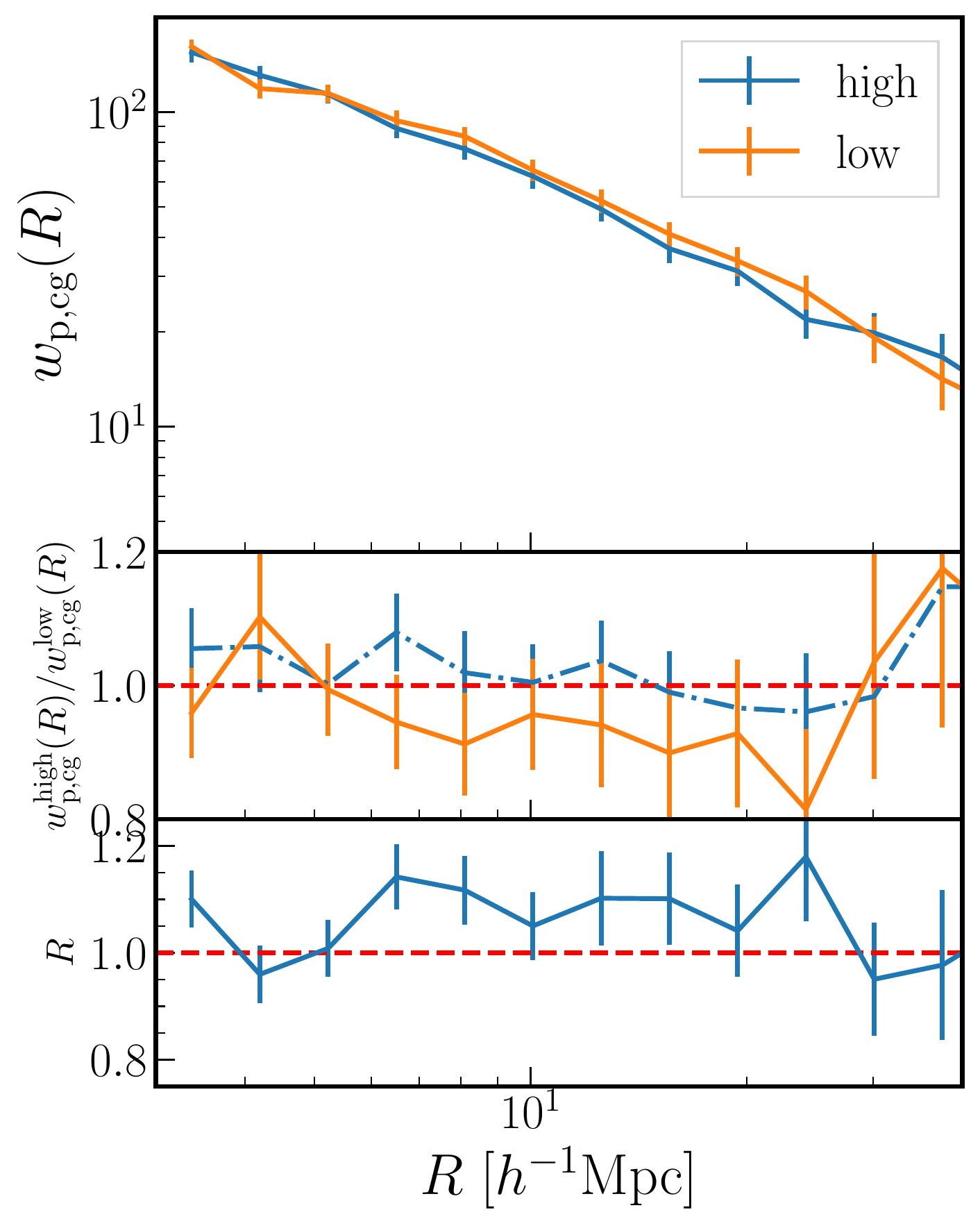}
\caption{\label{fig:ror_mstar} Top: Cluster-galaxy projected cross-correlation functions integrated up to $\pi_{\rm max}=100h^{-1}{\rm Mpc}$ for the high/low-$M_*$ subsamples. Middle:The ratio $w_{\rm p,cg}^{\rm high}(R)$ and $w_{\rm p,cg}^{\rm low}(R)$ computed with $\pi_{\rm max}=50h^{-1}{\rm Mpc}$ (blue dash-dot) and $\pi_{\rm max}=100h^{-1}{\rm Mpc}$ (orange solid). Bottom: The ratio of $w_{\rm p,cg}^{\rm high}(R)/w_{\rm p,cg}^{\rm low}(R)$ at $\pi_{\rm max}=50h^{-1}{\rm Mpc}$ and $\pi_{\rm max}=100h^{-1}{\rm Mpc}$ denoted as "Ratio of Ratio".}
\end{figure}

\begin{figure}[h!]
\centering
\includegraphics[width=0.85\textwidth]{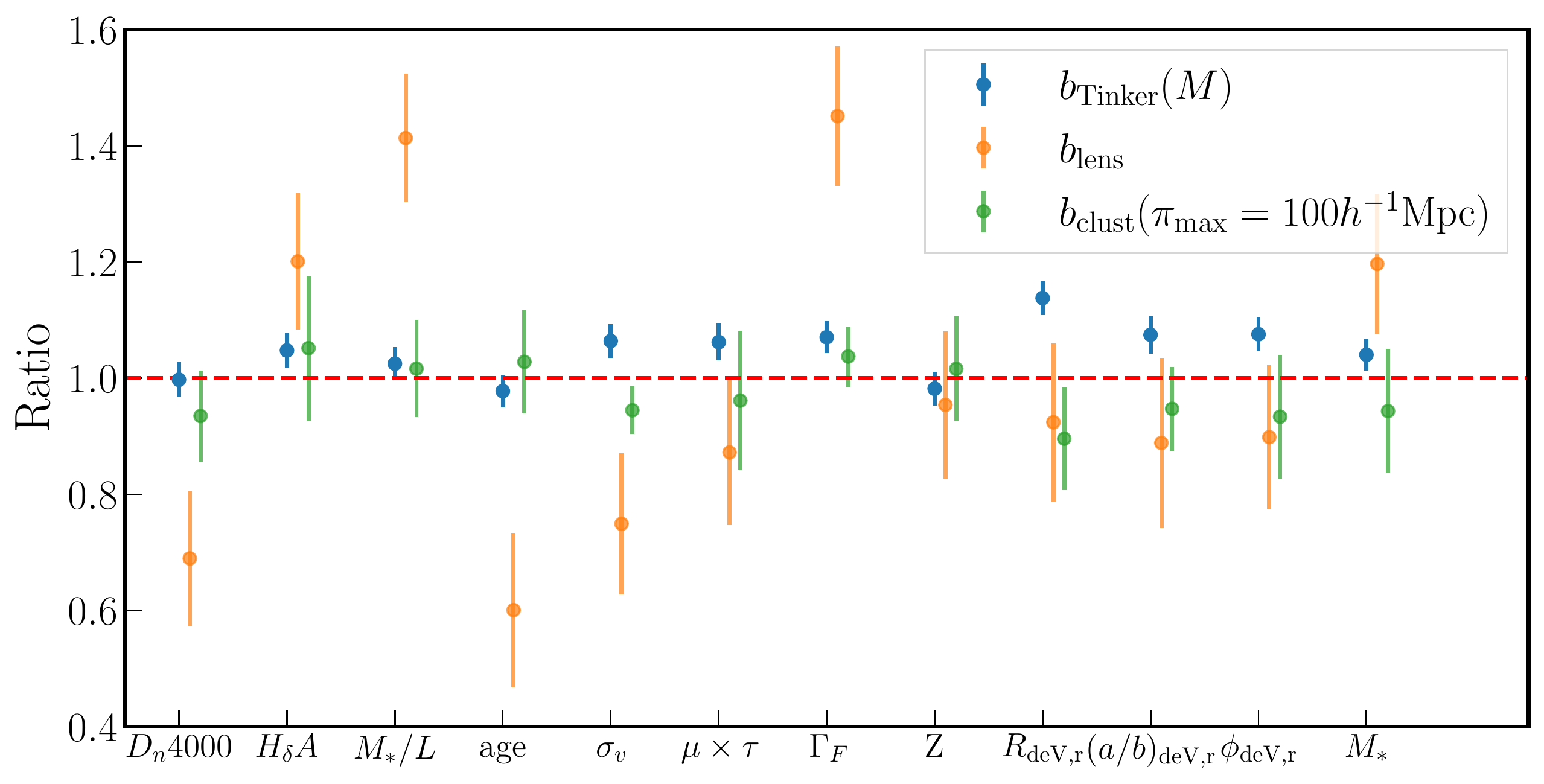}
\caption{\label{fig:bias}Bias ratio for the subsamples split by the central galaxy properties. $b_{\rm Tinker}(M)$ is the predicted halo bias based on halo mass using the Tinker bias model. The symbols $b_{\rm lens}$ and $b_{\rm clust}$ are the halo biases measured from lensing and clustering, respectively. For the bias ratio measured from clustering, we use $\pi_{\rm max}=100\mpch$.}
\end{figure}

\begin{figure}[h!]
\centering
\includegraphics[width=0.8\textwidth]{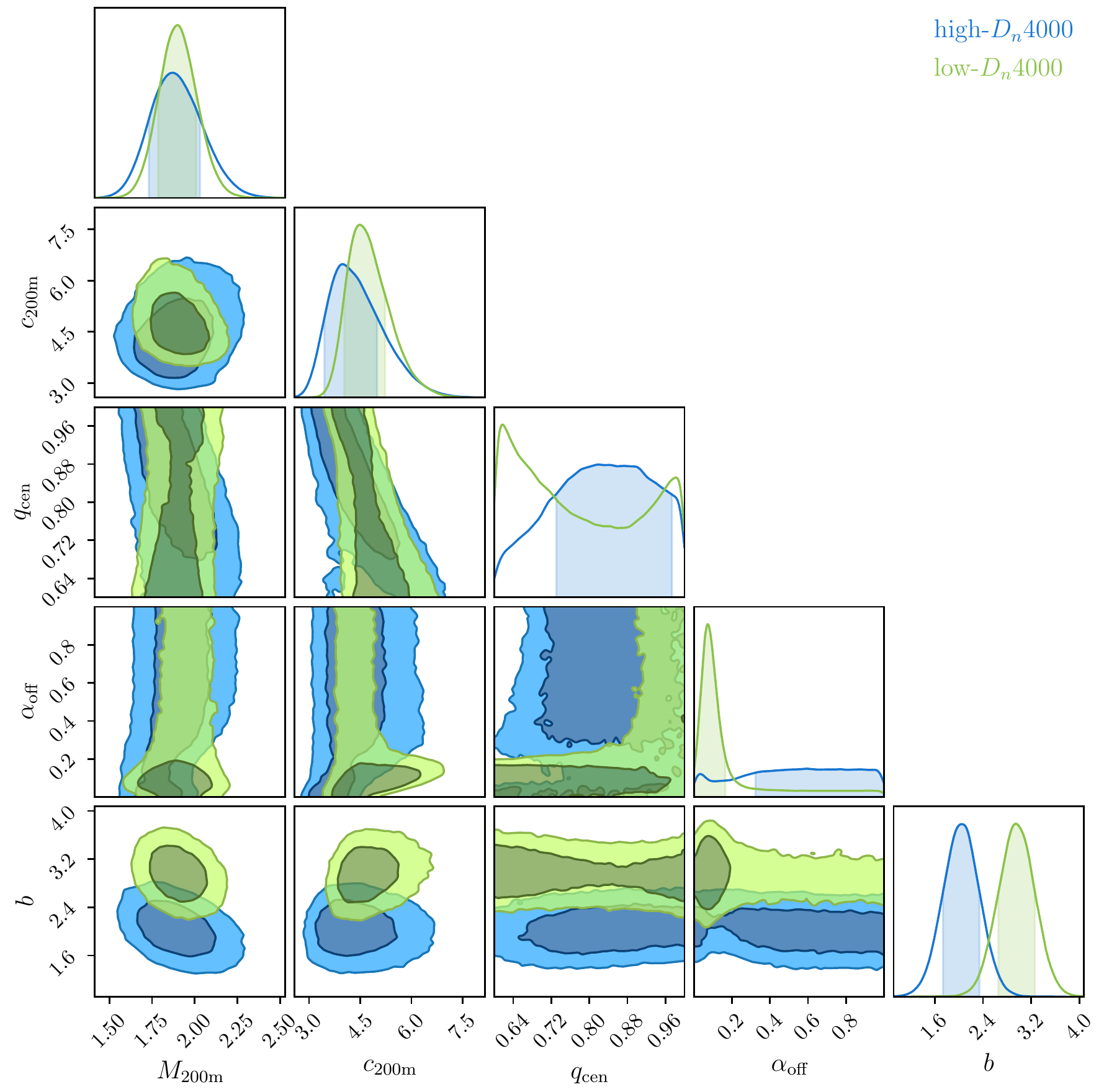}
\caption{\label{fig:triangle_dn4000}  Posterior constraint from the modelling of $\Delta \Sigma(R)$ for the high-$D_n 4000$ (blue) and low-$D_n 4000$ (green) subsamples. Each histogram in the diagonal panels are the 1D marginalised posterior distribution of the parameters, while each contour plot in the off-diagonal panels indicate the 1$\sigma$ and 2$\sigma$ confidence regions of the matching parameter pairs.}
\end{figure}